\documentclass[twocolumn,showpacs,preprintnumbers,amsmath,amssymb,floatfix]{revtex4}
\usepackage{graphicx}
\usepackage{slashed}
\usepackage{color}
\begin{document}

\title{Role of the high-spin nucleon and delta resonances in the 
  $K\Lambda$ and $K\Sigma$  photoproduction off the nucleon}

\author{N. H. Luthfiyah}
\author{T. Mart}
\email[Corresponding author: ]{terry.mart@sci.ui.ac.id}

\affiliation{Departemen Fisika, FMIPA, Universitas Indonesia, Depok 16424, Indonesia}


\begin{abstract}
We have investigated the effect of  nucleon and delta resonances with spins 11/2,
13/2, and 15/2 in the kaon photoproduction process $\gamma + N \to K + Y$ 
by using two covariant isobar models. The formalism for high-spin propagators and 
interaction Lagrangians were adopted from the works of Pascalutsa and Vrancx {\it et al}. 
The calculated scattering amplitudes were decomposed into six Lorentz- and gauge-invariant 
matrices, from which we calculated the cross sections and polarization observables. 
The unknown parameters in the amplitudes, i.e., the coupling constants and hadronic 
form factor cutoffs, were obtained by fitting the calculated observables to experimental 
data. In the $K\Lambda$ channels the inclusion of $N(2600)I_{1,11}$ 
and $N(2700)K_{1,13}$ resonances improves the agreement between model calculations and
experimental data significantly and reduces the dominance of resonances in the model by
increasing the hadronic form factor cutoff of the Born terms. Furthermore, the inclusion
of these resonances reduces the number of resonance structures in cross sections, including
the structure in the $K^0\Lambda$ differential cross section 
at $W\approx 1650$ MeV, which could become a hint of the narrow resonance.
In the $K\Sigma$ channels the inclusion of 
$N(2600)I_{1,11}$, $N(2700)K_{1,13}$, $\Delta(2420)H_{3,11}$, $\Delta(2750)I_{3,13}$, and
$\Delta(2950)K_{3,15}$ states also significantly improves the model and increases the 
hadronic form factor cutoff of the Born terms. However, different from the $K\Lambda$ channels, 
the inclusion of these high-spin resonances leads to more resonance structures in the 
$K^+\Sigma^0$  differential cross section. 
This investigation reveals that the second and third
peaks in the $K^+\Sigma^0$ differential cross section originate from the $\Delta(2000)F_{35}$
and $N(2290)G_{19}$ resonances, respectively.
We have also evaluated the resonance properties at the pole positions and using the Breit-Wigner
method. In both $K\Lambda$ and $K\Sigma$ channels the inclusion of the high-spin baryon 
resonances was found to improve the agreement between the resonance properties obtained 
in this study and those listed by the Particle Data Group.  
\end{abstract}


\maketitle
\section{INTRODUCTION}
\begin{figure*}[!t]
  \centering
  \includegraphics[scale=0.6]{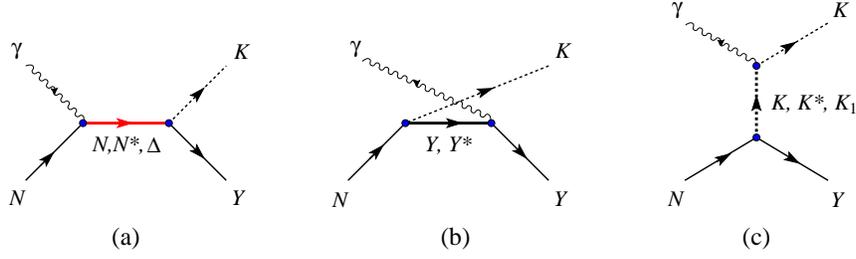}
  \caption{Feynman diagrams of the kaon photoproduction $\gamma(k) + N(p) \to K(q) + Y (p_Y)$
    for the 
    (a) $s$-channel, (b) $u$-channel, and (c) $t$-channel intermediate states.}
  \label{fig:feynman}
\end{figure*}

Recently, the effect of spins-7/2 and -9/2  nucleon resonances on kaon photoproduction 
processes has been phenomenologically  investigated by using a covariant isobar model
~\cite{Clymton:2017nvp}, in which the scattering amplitude was calculated by using 
the appropriate Feynman diagrams depicted in Fig.~\ref{fig:feynman}. The analytical 
calculation performed in this study made use of the consistent interaction Lagrangians 
proposed by Pascalutsa~\cite{Pascalutsa:2000kd}. The calculated observables were fitted 
to nearly 7400 experimental data points. The result of the fitting process showed that the 
inclusion of spins-7/2 and 9/2 nucleon resonances could improve the agreement between the 
model calculation and the experimental data. The model was later extended to describe 
both $\gamma p \to K^+ \Lambda $ and $\gamma n \to K^0 \Lambda $ processes, simultaneously
\cite{Mart:2019mtq}. In the latter, the model was fitted to nearly 9400 data points, 
including recent data from the CLAS and MAMI collaborations. The extended model yielded
a nice agreement between the calculated observables and experimental data in both 
isospin channels. 

Despite the success of the model, it has not yet consider the resonances with 
spins-11/2, 13/2 and 15/2, which are tabulated by the Particle Data Group (PDG) 
listing \cite{pdg} (see Table~\ref{tab:resonance}). Given the fact 
that the inclusion of spins-7/2 and 9/2 resonances in Ref.~\cite{Clymton:2017nvp} 
significantly improves the model, we could also expect that a similar 
phenomenon would be obtained with the inclusion of  the resonances  listed in 
Table~\ref{tab:resonance}. Furthermore, we can also extend the model to include
the four $K\Sigma$ isospin channels, i.e. $K^+ \Sigma^0$, $K^0 \Sigma^+$, 
$K^+ \Sigma^-$ and $K^0 \Sigma^0$ channels. Since the total isospin of these channels 
is 3/2, the $\Delta$ resonances are allowed as the intermediate states. An isobar 
model for $K\Sigma$ reaction can be constructed from the well-known $K\Lambda$ model 
based on our previous studies, e.g., in Ref.~\cite{Clymton:2019nrs}. For the $K\Lambda$ 
final states, the isospin symmetry couples the $K^+ \Lambda$ and $K^0 \Lambda$ channels, 
whereas for the $K\Sigma$ final states, the $K^+\Sigma^0$, $K^0\Sigma^+$, $K^+\Sigma^-$, 
and $K^0\Sigma^0$ production processes are coupled by isospin symmetry to one model. 
As mentioned above, there are nearly 9400 data points available for the $K\Lambda$ 
channels, and nearly 8000 data points for $K\Sigma$ ones. These data sets will be 
fitted to the $K\Lambda$ and $K\Sigma$ models separately. In principle, all six 
isospin channels can also be coupled, since all $K\Lambda$ and $K\Sigma$ isospin channels 
utilize the same leading Born coupling constants, i.e., the $g_{K \Lambda N}$
and $g_{K \Sigma N}$. However, since we have fixed these coupling constants to
the SU(3) values \cite{Adelseck:1990ch}, the $K\Lambda$ and $K\Sigma$ channels are
naturally decoupled.

\begin{table}[!b]
  \begin{center}
    \caption{Nucleon and delta resonances with spins from 11/2 to 15/2 used 
      in our study and tabulated by the PDG in their 
      particle listing~\cite{pdg}.}
    \label{tab:resonance}
    \begin{ruledtabular}
      \begin{tabular}{l c c c c c}
        Resonance & $J^P$ & Status & Mass (MeV) & Width (MeV) \\
        \hline
        $N(2600)I_{1,11}$& $11/2^-$ & *** & $2600\pm50$ & $650\pm150$ \\
        $N(2700)K_{1,13}$& $13/2^+$ & ** & $2612\pm45$ & $350\pm50$ \\
        $\Delta(2420)H_{3,11}$& $11/2^+$ & **** & $2450\pm150$ & $500\pm200$ \\
        $\Delta(2750)I_{3,13}$& $13/2^-$ & ** & $2794\pm80$ & $350\pm100$ \\
        $\Delta(2950)K_{3,15}$& $15/2^+$ & ** & $2990\pm100$ & $330\pm100$ \\
      \end{tabular}
    \end{ruledtabular}
  \end{center}
\end{table} 

To the best of our knowledge, there are limited studies investigating the contribution 
of high-spin nucleon and delta resonances in kaon photoproduction with the field-theoretic 
model. Presumably, this is due to the complicated formulations of propagator and vertex 
factors along with the problem of lower-spin background that plagued the formulation 
of high-spin ($J>1/2$) resonance propagator. Therefore, the first purpose of this paper is 
to set forth the formulation of higher-spin resonances amplitude. After that we can study 
their effect on the six isospin channels of kaon photoproduction (see Table~\ref{tab:channel})
by means of a covariant isobar model.
\begin{table}[t]
  \begin{center}
    \caption{Six possible isospin channels of kaon photoproduction on the nucleon along with
    their threshold energies in terms of photon lab energy $k^{\rm thr.}$ and total c.m.
    energy $W^{\rm thr.}$.}
    \label{tab:channel}
    \begin{ruledtabular}
      \begin{tabular}{c l c c}
        \multicolumn{ 1}{c}{No.} & \multicolumn{ 1}{l}{Reactions} & \multicolumn{ 1}{c}{\,\,\,$k^{\rm{thr.}}$ (MeV)\,\,\,} & \multicolumn{ 1}{c}{\,\,\,$W^{\rm{thr.}}$ (MeV)\,\,\,} \\
        \hline
        1) & $\gamma + p \to K^+ + \Lambda$\,\,\,\,\,\,\,\, & 911 & 1609 \\
        2) & $\gamma + n \to K^0 + \Lambda$\,\,\,\,\,\,\,\, & 915 & 1613 \\
        3) & $\gamma + p \to K^+ + \Sigma^0$\,\,\,\,\,\,\,\, & 1046 & 1686 \\
        4) & $\gamma + p \to K^0 + \Sigma^+$\,\,\,\,\,\,\,\, & 1048 & 1687 \\
        5) & $\gamma + n \to K^+ + \Sigma^-$\,\,\,\,\,\,\,\, & 1052 & 1691 \\
        6) & $\gamma + n \to K^0 + \Sigma^0$\,\,\,\,\,\,\,\, & 1051 & 1690 \\
      \end{tabular}
    \end{ruledtabular}
  \end{center}
\end{table} 

We have organized this paper as follows. In Sec.~\ref{sec:formalism} we present the formalism 
used in our study. In Sec.~\ref{sec:result} we present the numerical result and discuss
the comparison between model calculations and experimental data. Finally, in Sec.~\ref{sec:summary} 
we summarize and conclude our work. The extracted form functions used to calculate the observables
for the fitting process are given in Appendix 
\ref{app:form_function}. 

\section{Formalism}
\label{sec:formalism}
As mentioned above we adopt the formalism of the nucleon propagators and the
interaction Lagrangian developed by Pascalutsa \cite{Pascalutsa:2000kd} and 
Vrancx \textit{et al.} \cite{Vrancx:2011qv}. In our previous work, we
explained this formalism in details and constructed the reaction amplitude 
for the nucleon resonances with spins 7/2 and 9/2 \cite{Clymton:2017nvp}. 
To facilitate the reader, in this section we briefly discuss this formalism 
and derive the construction of the spins-11/2, -13/2, and -15/2
resonance propagators along with their interaction Lagrangians. A 
preliminary result for the analytical form of the production amplitudes 
involving nucleon resonances with spins up to 13/2 has been reported in 
a conference~\cite{Luthfiyah:2020ogb}. The notation of the four-momentum of
photon, nucleon, kaon, and hyperon used in the following discussion is
given in the caption of Fig.~\ref{fig:feynman}, with $p_R=p+k=p_Y+q$.

\subsection{Consistent Interaction Theory}
A consistent interaction Lagrangian is required to eliminate the appearance 
of the lower-spin background amplitude, which is known as an intrinsic problem
in the Rarita-Schwinger (R-S) formulation of the spin-3/2 (or higher) propagator. 
A number of solutions have been put forward to solve this problem in the last 
decades. Among them, those of Pascalutsa~\cite{Pascalutsa:2000kd} and 
Vrancx {\it et al.}~\cite{Vrancx:2011qv} are relevant to our present work. They constructed
the interaction Lagrangians which automatically cancel out the lower spin contributions
to the scattering amplitude. Pascalutsa proposed a gauge-invariant 
interaction structure for spin-3/2 particles by introducing a local symmetry 
to the R-S field $\psi_{\mu_1...\mu_n}$. 
The local symmetry reads \cite{Vrancx:2011qv}
\begin{eqnarray}\label{eq:trans}
\psi_{\mu_1...\mu_n} \rightarrow \psi_{\mu_1...\mu_n}\, + \frac{1}{n(n -1)!}\, 
\sum_{P(\mu)}\, \partial_{\mu_1}\, \xi_{\mu_2...\mu_n},
\end{eqnarray}
where the totally symmetric and space-time dependent tensor-spinor field 
$\xi_{\mu_2...\mu_n}$ fulfills
\begin{eqnarray}
\gamma^{\mu_1}\, \xi_{\mu_1\mu_2...\mu_{n-1}} = 0 ~.
\end{eqnarray}

To construct a consistent interaction Lagrangian for particles with spin-3/2 we
need a gauge-invariant field $G_{\mu\nu}$, taken from Ref.~\cite{Pascalutsa:2000kd}, 
and the interaction operator  $O^{(3/2)}_{(\mu,\nu)\lambda}$, adopted from 
Ref.~\cite{Vrancx:2011qv}. The invariant field and the interaction operator 
is written in the form of
\begin{eqnarray}\label{eq:medan}
G_{\mu\nu} &=& \partial_\mu \psi_\nu \, -\, \partial_\nu \psi_\mu \nonumber \\
&=& (\partial_\mu g_{\nu\lambda} - \partial_\nu g_{\mu\lambda})\psi^\lambda \nonumber \\ 
&=& O^{(3/2)}_{(\mu,\nu)\lambda}\psi^\lambda ,
\end{eqnarray}
with
\begin{equation}
O^{(3/2)}_{(\mu,\nu)\lambda} = (\partial_\mu g_{\nu\lambda} - \partial_\nu g_{\mu\lambda}) , 
\end{equation}
where $\psi_\mu$ is the massive R-S field that obeys the R-S equation and its constraints. 
The interaction operator fulfills the following property,
\begin{eqnarray}
\partial^\lambda \, O^{3/2}_{(\mu,\nu)\lambda}\, (\partial) = \, O^{3/2}_{(\mu,\nu)\lambda}\, 
(\partial)\,\partial^\lambda ,
\end{eqnarray} 
which maintains the invariance of $G_{\mu\nu}$ under  the uR-$\rm{S_{5/2}}$ gauge as stated 
in Eq.~(\ref{eq:trans}).  With the invariant field and spin-3/2 interaction operator, 
Vrancx  {\it et al.} constructed a consistent interaction theory for spin-5/2 field. 
This formulation is later expanded for general high-spin field. Vrancx  {\it et al.} 
defined a general interaction operator as \cite{Vrancx:2011qv}
\begin{eqnarray}
&O^{(n + 1/2)}_{(\mu_1...\mu_n,\nu_1...\nu_n)\lambda_1...\lambda_n}\,(\partial) = &\nonumber\\
&{\displaystyle \frac{1}{(n!)^2}\sum_{P(\nu)}\sum_{P(\lambda)}}O^{(3/2)}_{(\mu_1,\nu_1)\lambda_1}\,
(\partial)\,...\,O^{(3/2)}_{(\mu_n,\nu_n)\lambda_n}\,(\partial)\, ,&
\end{eqnarray}
with $P(\mu)$$P(\lambda)$ indicate all possible permutations for all  $\mu$ and $\lambda$.
Vrancx reduced the indices of the $G_{\mu\nu}$ field introduced by Pascalutsa in 
Eq.~(\ref{eq:medan}), because the $G_{\mu\nu}$ field contains too many indices 
compared to the original field $\psi_{\mu_1..\mu_n}$. A new field is introduced 
by Vrancx as $\Psi_{\mu} = O^{(3/2)}_{(\mu,\nu)\lambda}\,\psi^\lambda\,\gamma^\nu$. 
For particles with spin-$(n + 1/2)$ the gauge-invariant field can be written as 
\begin{eqnarray}
\Psi_{\mu_1...\mu_n} =O^{(n + 1/2)}_{(\mu_1...\mu_n,\nu_1...\nu_n)\lambda_1...\lambda_n}\,
(\partial)\,\psi^{\lambda_1...\lambda_n}\,\gamma^{\nu_1...\nu_n} .
\end{eqnarray}
The interaction operator for the above field reads
\begin{eqnarray}\label{eq:oin}
&\mathcal{O}^{(n + 1/2)}_{(\mu_1...\mu_n)\lambda_1...\lambda_n}\,(\partial) = 
\gamma^{\nu_1..\nu_n}O^{(n + 1/2)}_{(\mu_1...\mu_n,\nu_1...\nu_n)\lambda_1...\lambda_n}\,
(\partial)& \nonumber \\
&= {\displaystyle \frac{1}{(n!)^2}\sum_{P(\lambda)}}\mathcal{O}^{(3/2)}_{(\mu_1)\lambda_1}\,
(\partial)\,...\,\mathcal{O}^{(3/2)}_{(\mu_n)\lambda_n}\,(\partial) \, ,&
\end{eqnarray} 
where
\begin{eqnarray}
\mathcal{O}^{(3/2)}_{(\mu_n)\lambda_n}\,(\partial)  = \gamma^\nu \, O^{(3/2)}_{(\mu,\nu)
\lambda}\,(\partial) \, .
\end{eqnarray}
With this consistent interaction we are ready to construct 
the Lagrangians for high-spin interactions.

\subsection{Interaction Lagrangians}
The basic Lagrangian for the kaon-hyperon-nucleon interaction is 
\begin{eqnarray}
\mathcal{L}_\mathrm{had} = g_{KYN}\,\bar{\Psi}_Y\,\gamma_5\,\Psi_N\,\Phi_K .
\end{eqnarray}
Following this basic Lagrangian, the standard interaction Lagrangian for the interaction 
with spin-1/2 resonances is written as
\begin{eqnarray}
\mathcal{L}_\mathrm{had} = g_{KYR}\,\bar{\Psi}_Y\,\gamma_5\,\Psi_{R}\,\Phi_K + {\rm H.c.}
\end{eqnarray}
According to Pascalutsa the hadronic interaction Lagrangian for spin-3/2 resonance
with mass $m_R$ reads \cite{Pascalutsa-PRD1998}
\begin{eqnarray}
\mathcal{L}_\mathrm{had}=\frac{g_{KYR}}{m_{R}^2}\, \epsilon^{\mu\nu\alpha\beta}\, 
\bar{\Psi}_Y \, \partial_\beta \phi^{*} \, \gamma_5 \, \gamma_\alpha \, (\partial_\mu 
\psi_\nu) + \mathrm{H.c} ,
\end{eqnarray}
where $\bar{\Psi}_Y$ is the spinor field of the hyperon,  $\phi$ is the pseudoscalar field 
of the kaon, and $\psi_\nu$ is the massive R-S field of the nucleon or resonance. 
However, the above Lagrangian is inconsistent for higher-spin resonances. With the 
substitution proposed by Vrancx, $\psi_\mu\, \rightarrow \, \bar{\Psi}_\mu / m_{R}$, 
the Lagrangian for a resonance with spin 3/2 can be written as
\begin{eqnarray}
\mathcal{L}_\mathrm{had}&=&\frac{g_{KYR}}{m_{R}^3}\, \epsilon^{\mu\nu\alpha\beta}\, \bar{\Psi}_Y \, \partial_\beta \phi^{*} \, \gamma_5 \, \gamma_\alpha \, \partial_\mu \Psi_\nu + \mathrm{H.c.} ,~~
\end{eqnarray}
whereas the Lagrangian for hadronic interaction with spin-$(n+1/2)$ resonances has the form of
\begin{eqnarray}\label{eq:lhad}
\mathcal{L}_\mathrm{had}&=&\frac{g_{KYR}}{m_{R}^{2n+1}}\, \epsilon^{\mu\nu_n\alpha\beta}\,
\partial^{\nu_1}\,...\partial^{\nu_{n-1}}\, \bar{\Psi}_Y \, \partial_\beta \phi^{*} \nonumber\\ 
&& \times \gamma_5 \, \gamma_\alpha \, (\partial_\mu \Psi_{\nu_1...\nu_n}) + \mathrm{H.c} \nonumber \\
&=& \frac{g_{KYR}}{m_{R}^{2n+1}}\, \epsilon^{\mu\nu\alpha\beta}\, \bar{\Psi}_Y \, 
\partial_\beta \phi^{*} \, \gamma_5 \, \gamma_\alpha \nonumber\\ 
&& \times\partial_\mu O^{(n + 1/2)}_{(\mu_1...\mu_n,\nu_1...\nu_n)\lambda_1...\lambda_n}\,
(\partial) \nonumber \\
& &\times \gamma^{\alpha_1}\,...\gamma^{\alpha_n}\, \psi^{\lambda_1...\lambda_n} + {\rm H.c}.
\end{eqnarray}
In addition to the hadronic Lagrangian, Pascalutsa has also constructed the Lagrangian for 
the electromagnetic interaction, which is written as \cite{Pascalutsa-PRC1999}
\begin{eqnarray}\label{eq:lag-em}
\mathcal{L}_{\rm em} &=& \frac{e}{m_{R}^3}\bar{\Psi}^\beta \Bigl\{(g_1\,\epsilon_{\mu\nu\alpha\beta} 
\partial^\alpha \Psi + g_2\,\gamma_5 \, g_{\beta\nu}\partial_\mu \Psi \nonumber\\
&&+ g_3\,\gamma_\mu\, \gamma^\rho\,\epsilon_{\rho\nu\alpha\beta}\partial^\alpha \Psi 
+ g_4\,\gamma_5\,\gamma_\mu\,\gamma^\rho \nonumber\\
& & \times (\partial_\rho g_{\nu\beta} - \partial_\nu g_{\rho\beta}) \Psi \Bigr\}F^{\mu\nu} 
+ \mathrm{H.c} \, ,
\end{eqnarray}
where $F^{\mu\nu}$ is the conventional electromagnetic field strength tensor.
To correctly model the interaction in discussion, a consistent interaction for the electromagnetic 
interaction Lagrangian is needed. The electromagnetic interaction Lagrangian for 
the resonance particles with spin-$(n + 1/2)$ can be written as
\begin{eqnarray}\label{eq:lem}
\mathcal{L}_{\rm em} &=& \frac{e}{m_{R}^{2n+1}}\bar{\Psi}^{\beta_1\cdots\beta_n}\bigl\{(g_1\,\epsilon_{\mu\nu\alpha\beta_n} \partial^\alpha \Psi \nonumber\\
& & + g_2\,\gamma_5 \, g_{\beta_n\nu}\partial_\mu \Psi + g_3\,\gamma_\mu\,\gamma^\rho\,\epsilon_{\rho\nu\alpha\beta_n}\partial^\alpha \Psi  \nonumber\\
& & + g_4\,\gamma_5\,\gamma_\mu\,\gamma^\rho\,(\partial_\rho g_{\nu\beta_n} - \partial_\nu g_{\rho\beta_n}) \Psi \bigr\} \nonumber\\
&& \times \partial_{\beta_1} \cdots\partial_{\beta_{n-1}} F^{\mu\nu} + \mathrm{H.c.}
\end{eqnarray}

The consistency of the interaction is guaranteed by the operator interaction which fulfills:
\begin{eqnarray}\label{eq:sifatopint}
p_{R}^{\lambda_i}\, O^{(n + 1/2)}_{(\mu_1...\mu_n,\nu_1...\nu_n)\lambda_1...\lambda_n} (p_{R}) = 0 ,
\end{eqnarray}
with $i = 1,2,...,n$ and $p_{R}^{\lambda_i}$ is the four-momenta of the resonance particles with spin-$(n + 1/2)$. 

The electromagnetic and hadronic vertices, which are required to calculate the scattering amplitude,
can be obtained from the interaction Lagrangians constructed in the previous discussion. 
As the result the hadronic vertex can be written as
 \begin{eqnarray}
 \Gamma^\mathrm{had}_{\mu_1...\mu_2} &=& \frac{g_{KYR}}{m_{R}^{2n+1}}\,\epsilon^{\mu\nu_n\alpha\beta}\,p_{\Lambda}^{\nu_1}\,...p_{\Lambda}^{\nu_{n-1}}\,q_\beta\,\gamma_5\,\gamma_\alpha\,{p_R}_\mu \nonumber \\
 & & \times O^{(n + 1/2)}_{(\nu_1...\nu_n,\alpha_1...\alpha_n)\mu_1...\mu_n}\,(p_{R}) \gamma^{\alpha_1}\,...\gamma^{\alpha_n} . ~~~
 \end{eqnarray}
and the electromagnetic vertex reads 
 \begin{eqnarray}
 \Gamma^{\rm em}_{\nu_1\cdots\nu_n} &=& \frac{e}{m_{R}^{2n+1}}\,O_{n+1/2}^{(\beta_1 \cdots \beta_n , 
   \alpha_1 \cdots \alpha_n)\nu_1 \cdots \nu_n}(p_R)\gamma_{\alpha_1} \cdots \gamma_{\alpha_n} \nonumber\\
 && \times \,\Bigl\{g_1\,\epsilon_{\mu\nu\alpha\beta_n} p^\alpha + g_2\,\gamma_5 \, g_{\beta_n\nu}p_\mu \nonumber\\
 & & + g_3\,\gamma_\mu\,\gamma^\rho\,\epsilon_{\rho\nu\alpha\beta_n} p^\alpha \nonumber\\
&& + g_4\,\gamma_5\,\gamma_\mu\,\gamma^\rho\,(p_\rho g_{\nu\beta_n} - p_\nu g_{\rho\beta_n})\Bigr\}\nonumber\\
 & & \times k_{\beta_1} \cdots k_{\beta_{n-1}}\,(k^\mu \epsilon^\nu - k^\nu \epsilon^\mu) + \mathrm{H.c.}
 \end{eqnarray}
 The vertex factors can be simplified to 
\begin{eqnarray} \label{eq:verhad}
  \Gamma^\mathrm{had}_{\mu_1...\mu_2} &=& \frac{1}{m_{R}^n} \tilde{\Gamma}_\mathrm{had}^{\nu_1...\nu_n}\,\tilde{O}^{(n + 1/2)}_{(\nu_1...\nu_n)\mu_1...\mu_n}\,(p_{R})
\end{eqnarray}
 \begin{eqnarray}\label{eq:verem}
 \Gamma_\mathrm{em}^{\mu_1...\mu_2} &=& \frac{1}{m_{R}^n}\,\tilde{O}^{(n + 1/2)}_{(\nu_1...\nu_n)\mu_1...\mu_n}\,(p_{R})\, \tilde{\Gamma}^\mathrm{em}_{\nu_1...\nu_n} 
 \end{eqnarray}
where $\tilde{O}^{(n + 1/2)}_{(\nu_1...\nu_n)\mu_1...\mu_n}\,(p_{R})$ is the interaction operator defined by
\begin{eqnarray}\label{eq:intop}
  &\tilde{O}^{(n + 1/2)}_{(\nu_1...\nu_n)\mu_1...\mu_n}\,(p_{R}) = &\nonumber\\
&O^{(n + 1/2)}_{(\nu_1...\nu_n,\alpha_1...\alpha_n)\mu_1...\mu_n}\,(p_{R}) \gamma^{\alpha_1}\,...\gamma^{\alpha_n} ~.&
 \end{eqnarray}

In the present work we use the propagator 
\begin{eqnarray}\label{eq:propag}
&P_{\mu_1..\mu_n;\nu_1..\nu_n}\,(p_R) = &\nonumber\\
&{\displaystyle \frac{\slashed{p}_R + m_R}{p_R^2 - m_R^2 + im_R\Gamma}}\,
\tilde{\mathcal{P}}^{n+1/2}_{\mu_1...\mu_n;\nu_1...\nu_n}(p_R) , &
\end{eqnarray}
where  $\tilde{\mathcal{P}}^{n+1/2}_{\mu_1...\mu_n;\nu_1...\nu_n}(p_R)$ is the on-shell projection operator. 
The complicated form of the projection operator will be discussed later.

The production amplitude is obtained by sandwiching the propagator between the two vertices, i.e.,
\begin{eqnarray}
\label{eq:amplitude_initial}
\mathcal{M}_{\mathrm{res}}^\mathrm{n+1/2}= \bar{u}_\Lambda \Gamma^{\rm had}_{\mu_1...\mu_n}P^{\mu_1...\mu_n,\nu_1...\nu_n}_{(n+1/2)}\,(p_{R})\,\Gamma^{\rm em}_{\nu_1...\nu_n}\,u_p .~
\end{eqnarray}
By inserting Eqs.~(\ref{eq:verhad}), (\ref{eq:verem}), and (\ref{eq:propag}) into 
Eq.~(\ref{eq:amplitude_initial}) we obtain
\begin{eqnarray}
\label{eq:mres}
\mathcal{M}_{\mathrm{res}}^\mathrm{n+1/2}&=& \bar{u}_\Lambda\tilde{\Gamma}^{\alpha_1...\alpha_n}_{\rm had}\tilde{O}^{n + 1/2}_{(\alpha_1...\alpha_n)\mu_1...\mu_n}\,(p_{R})\nonumber\\
&&\times \frac{1}{m_{R}^{2n}}\,P^{\mu_1...\mu_n,\nu_1...\nu_n}_{(n+1/2)}\,(p_{R})\nonumber \\
&& \times \tilde{O}_{n + 1/2}^{(\beta_1...\beta_n)\nu_1...\nu_n}\,(p_{R})\,\tilde{\Gamma}_{\rm em}^{\beta_1...\beta_n}\,u_p ,
\nonumber\\
&=& \bar{u}_\Lambda\tilde{\Gamma}^{\alpha_1...\alpha_n}_{\rm had}\tilde{O}^{n + 1/2}_{(\alpha_1...\alpha_n)\mu_1...\mu_n}\,(p_{R})\nonumber\\
&& \times \frac{1}{m_{R}^{2n}}\,\frac{\slashed{p}_{R} + m_{R}}{p_{R}^2 - m_{R}^2 + im_{R}\Gamma }\nonumber  \; \\&&\times \mathcal{P}^{\mu_1...\mu_n,\nu_1...\nu_n}_{(n+1/2)}\,(p_{R}) \tilde{O}_{n + 1/2}^{(\beta_1...\beta_n)\nu_1...\nu_n}\,(p_{R})\nonumber\\
&& \times\tilde{\Gamma}_{\rm em}^{\beta_1...\beta_n}\,u_p ~.
\end{eqnarray}
The above formulation is simplified by considering the orthogonalities of the projection operator,
\begin{eqnarray}\label{eq:ortho}
\gamma_{\mu_i}\,\mathcal{P}^{\mu_1..\mu_n,\nu_1..\nu_n}_{(n + 1/2)}\,(p_{R}) &=& 
\mathcal{P}^{\mu_1..\mu_n,\nu_1..\nu_n}_{(n + 1/2)}\,(p_{R})\,\gamma_{\nu_i}\nonumber\\ &=& 0,
\end{eqnarray}
and 
\begin{eqnarray}
{p_R}_{\mu_i}\,\mathcal{P}^{\mu_1..\mu_n,\nu_1..\nu_n}_{(n + 1/2)}\,(p_{R}) &=& 
{p_R}_{\nu_i}\,\mathcal{P}^{\mu_1..\mu_n,\nu_1..\nu_n}_{(n + 1/2)}\,(p_{R})\nonumber \\
& =& 0 ,
\end{eqnarray}
with  $i = 1,2,..n $.

Thus, the scattering amplitude can be written as
\begin{widetext}
\begin{eqnarray}\label{scatt}
\mathcal{M}_{\mathrm{res}}^\mathrm{n+1/2} &=&\bar{u}_\Lambda\, \prod_{i=1}^{n}\slashed{p}_{R}\,g_{\alpha_i\,\mu_i}\tilde{\Gamma}_{\rm had}^{\alpha_1...\alpha_n}\,\frac{1}{m_{R}^{2n}}\,\frac{\slashed{p}_{R} + m_{R}}{p_{R}^2 - m_{R}^2 + im_{R}\Gamma } 
 \mathcal{P}^{\mu_1...\mu_n,\nu_1...\nu_n}_{(n+1/2)}\,(p_{R})\,\prod_{i=1}^{n}\slashed{p}_{R}\,g_{\beta_j\,\nu_j}\tilde{\Gamma}_{\rm em}^{\beta_1..
	\beta_n}\,u_p \nonumber \\
&=&\bar{u}_\Lambda\,\tilde{\Gamma}^{\rm had}_{\mu_1...\mu_n}\,\frac{p_R^{2n}}{m_{R}^{2n}}\,\frac{\slashed{p}_{R} + m_{R}}{p_{R}^2 - m_{R}^2 + im_{R}\Gamma}\,\mathcal{P}^{\mu_1...\mu_n,\nu_1...\nu_n}_{(n+1/2)}\,(p_{R})\,\tilde{\Gamma}_{\rm em}^{\beta_1..
	\beta_n}\,u_p \nonumber .\\
\end{eqnarray}
The above equation shows how a consistent interaction structure constructed intuitively. 

\subsection{Propagators}
The propagators used in this model are constructed from the corresponding particle 
projection operators. The generalized projection operator has been explored by 
Huang, \textit{et al.} in Ref.~\cite{Huang:2004we}. The projection operator for 
spin-11/2 can be written as
\begin{eqnarray}
\mathcal{P}^{11/2}_{\substack{\mu\mu_1\mu_2\mu_3\mu_4 \\ \nu\nu_1\nu_2\nu_3\nu_4}}
&=&
\frac{1}{1440} \sum_{\mathrm{P}(\mu),\mathrm{P}(\nu)}\Bigl\{ P_{\mu\nu}P_{\mu_1\nu_1}P_{\mu_2\nu_2}P_{\mu_3\nu_3}P_{\mu_4\nu_4}
-\textstyle{\frac{10}{11}}P_{\mu\mu_1}P_{\nu\nu_1}P_{\mu_2\nu_2}P_{\mu_3\nu_3}P_{\mu_4\nu_4} \nonumber\\
&&+\textstyle\frac{5}{33}P_{\mu\mu_1}P_{\nu\nu_1}P_{\mu_2\mu_3}P_{\nu_2\nu_3}P_{\mu_4\nu_4}    +\textstyle{\frac{5}{11}}\gamma^{\,\rho}\gamma^\sigma P_{\mu\rho}P_{\nu\sigma}P_{\mu_1\nu_1}P_{\mu_2\nu_2}P_{\mu_3\nu_3}P_{\mu_4\nu_4} \nonumber \\ & &
-\textstyle{\frac{10}{33}}\gamma^{\,\rho}\gamma^\sigma P_{\mu\rho}P_{\nu\sigma}P_{\mu_1\mu_2}P_{\nu_1\nu_2}P_{\mu_3\nu_3}P_{\mu_4\nu_4}  + \textstyle{\frac{5}{231}}\gamma^{\,\rho}\gamma^\sigma P_{\mu\rho}P_{\nu\sigma}P_{\mu_1\mu_2}P_{\nu_1\nu_2}P_{\mu_3\mu_4}P_{\nu_3\nu_4} \Bigr\}~,
\end{eqnarray}
and for spin-13/2 particles the projection operator is 
\begin{eqnarray}
\mathcal{P}^{13/2}_{\substack{\mu\mu_1\mu_2\mu_3\mu_4\mu_5 \\ \nu\nu_1\nu_2\nu_3\nu_4\nu_5}}
&=&
\frac{1}{(6!)^2}\sum_{\mathrm{P}(\mu),\mathrm{P}(\nu)}\Bigl\{ P_{\mu\nu}P_{\mu_1\nu_1}P_{\mu_2\nu_2}P_{\mu_3\nu_3}P_{\mu_4\nu_4}P_{\mu_5\nu_5} -\textstyle{\frac{15}{13}}P_{\mu\mu_1}P_{\nu\nu_1}P_{\mu_2\nu_2}P_{\mu_3\nu_3}P_{\mu_4\nu_4}P_{\mu_5\nu_5}\nonumber\\
& & +\textstyle\frac{45}{143}P_{\mu\mu_1}P_{\nu\nu_1}P_{\mu_2\mu_3}P_{\nu_2\nu_3}P_{\mu_4\nu_4}P_{\mu_5\nu_5} - \textstyle\frac{5}{429}P_{\mu\mu_1}P_{\nu\nu_1}P_{\mu_2\mu_3}P_{\nu_2\nu_3}P_{\mu_4\mu_5}P_{\nu_4\nu_5}\nonumber\\
& & +\textstyle{\frac{6}{13}}\gamma^{\,\rho}\gamma^\sigma P_{\mu\rho}P_{\nu\sigma}P_{\mu_1\nu_1}P_{\mu_2\nu_2}P_{\mu_3\nu_3}P_{\mu_4\nu_4}P_{\mu_5\nu_5}
-\textstyle{\frac{60}{143}}\gamma^{\,\rho}\gamma^\sigma P_{\mu\rho}P_{\nu\sigma}P_{\mu_1\mu_2}P_{\nu_1\nu_2}P_{\mu_3\nu_3}P_{\mu_4\nu_4}P_{\mu_5\nu_5} \nonumber\\
& & + \textstyle{\frac{30}{429}}\gamma^{\,\rho}\gamma^\sigma P_{\mu\rho}P_{\nu\sigma}P_{\mu_1\mu_2}P_{\nu_1\nu_2}P_{\mu_3\mu_4}P_{\nu_3\nu_4}P_{\mu_5\nu_5} \Bigr\}~,
\end{eqnarray}
while the projection operator for spin-15/2 particles reads
\begin{eqnarray}
\lefteqn{ 
\mathcal{P}^{15/2}_{\substack{\mu\mu_1\mu_2\mu_3\mu_4\mu_5\mu_6 \\ \nu\nu_1\nu_2\nu_3\nu_4\nu_5\nu_6}} = \frac{1}{(7!)^2}  \sum_{\mathrm{P}(\mu),\mathrm{P}(\nu)}\Bigl\{ P_{\mu\nu}P_{\mu_1\nu_1}P_{\mu_2\nu_2}P_{\mu_3\nu_3}P_{\mu_4\nu_4}P_{\mu_5\nu_5}P_{\mu_6\nu_6}} \nonumber\\ 
&& -\textstyle{\frac{21}{15}}P_{\mu\mu_1}P_{\nu\nu_1}P_{\mu_2\nu_2}P_{\mu_3\nu_3}P_{\mu_4\nu_4}P_{\mu_5\nu_5}P_{\mu_6\nu_6}+\textstyle\frac{7}{13}P_{\mu\mu_1}P_{\nu\nu_1}P_{\mu_2\mu_3}P_{\nu_2\nu_3}P_{\mu_4\nu_4}P_{\mu_5\nu_5}P_{\mu_6\nu_6} \nonumber\\
& & - \textstyle\frac{7}{143}P_{\mu\mu_1}P_{\nu\nu_1}P_{\mu_2\mu_3}P_{\nu_2\nu_3}P_{\mu_4\mu_5}P_{\nu_4\nu_5}P_{\mu_6\nu_6}
+\textstyle{\frac{7}{15}}\gamma^{\,\rho}\gamma^\sigma P_{\mu\rho}P_{\nu\sigma}P_{\mu_1\nu_1}P_{\mu_2\nu_2}P_{\mu_3\nu_3}P_{\mu_4\nu_4}P_{\mu_5\nu_5}P_{\mu_6\nu_6}\nonumber\\
& & -\textstyle{\frac{7}{13}}\gamma^{\,\rho}\gamma^\sigma P_{\mu\rho}P_{\nu\sigma}P_{\mu_1\mu_2}P_{\nu_1\nu_2}P_{\mu_3\nu_3}P_{\mu_4\nu_4}P_{\mu_5\nu_5}P_{\mu_6\nu_6} 
 + \textstyle{\frac{21}{143}}\gamma^{\,\rho}\gamma^\sigma P_{\mu\rho}P_{\nu\sigma}P_{\mu_1\mu_2}P_{\nu_1\nu_2}P_{\mu_3\mu_4}P_{\nu_3\nu_4}P_{\mu_5\nu_5}P_{\mu_6\nu_6} \nonumber \\
& & \textstyle{\frac{7}{1287}}\gamma^{\,\rho}\gamma^\sigma P_{\mu\rho}P_{\nu\sigma}P_{\mu_1\mu_2}P_{\nu_1\nu_2}P_{\mu_3\mu_4}P_{\nu_3\nu_4}P_{\mu_5\mu_6}P_{\nu_5\nu_6} \Bigr\}~,
\end{eqnarray}
The projection operators written above fulfill the orthogonality condition given 
by Eq.(\ref{eq:ortho}). According to Eq.(\ref{scatt}), the propagator for spin-11/2 
resonance can be written as
\begin{eqnarray}
\label{eq:prop_11}
P^{11/2}_{\substack{\mu\mu_1\mu_2\mu_3\mu_4 \\ \nu\nu_1\nu_2\nu_3\nu_4}}
&=& 
\frac{s^5}{m_{R}^{10}}\frac{(\slashed{p}+\slashed{k}+m_{R})}{(s-m_{R}^2+im_{R}\Gamma_{R})} \mathcal{P}^{11/2}_{\substack{\mu\mu_1\mu_2\mu_3\mu_4 \\ \nu\nu_1\nu_2\nu_3\nu_4}} \, ,
\end{eqnarray}
whereas for spin-13/2 and spin-15/2 resonances the propagators read
\begin{eqnarray}
\label{eq:prop_13}
P^{13/2}_{\substack{\mu\mu_1\mu_2\mu_3\mu_4\mu_5 \\ \nu\nu_1\nu_2\nu_3\nu_4\nu_5}} &=& 
\frac{s^6}{m_{R}^{12}}\frac{(\slashed{p}+\slashed{k}+m_{R})}{(s-m_{R}^2+im_{R}\Gamma_{R})} 
\mathcal{P}^{13/2}_{\substack{\mu\mu_1\mu_2\mu_3\mu_4\mu_5 \\ \nu\nu_1\nu_2\nu_3\nu_4\nu_5}}
\end{eqnarray}
and
\begin{eqnarray}
\label{eq:prop_15}
{P}^{15/2}_{\substack{\mu\mu_1\mu_2\mu_3\mu_4\mu_5\mu_6 \\ \nu\nu_1\nu_2\nu_3\nu_4\nu_5\nu_6}}
 &=& 
\frac{s^7}{m_{R}^{14}}\frac{(\slashed{p}+\slashed{k}+m_{R})}{(s-m_{R}^2+im_{R}\Gamma_{R})}\mathcal{P}^{15/2}_{\substack{\mu\mu_1\mu_2\mu_3\mu_4\mu_5\mu_6 \\ \nu\nu_1\nu_2\nu_3\nu_4\nu_5\nu_6}} \, ,
\end{eqnarray}
respectively, with $s = p_{R}^2 = (p + k)^2 = W^2$. Notice that the factors ${s^5}/{m_{R}^{10}}$, 
${s^6}/{m_{R}^{12}}$ and ${s^7}/{m_{R}^{14}}$ originate from the consequence of the 
consistent interaction. 

\subsection{Hadronic and Electromagnetic Vertices}

By using the above prescription we obtain that the hadronic and electromagnetic 
vertices for spin-11/2 reads as
\begin{eqnarray}
  \label{eq:had_11}
\Gamma_{\rm{had}}^{\mu\mu_1\mu_2\mu_3\mu_4\pm}&=& \frac{g_{KYR}}{m_{R}^6}\Gamma_{\mp}\bigl[(p_\Lambda\cdot q - \slashed{p}_\Lambda\slashed{q})\gamma^\mu + \slashed{p}_\Lambda q^{\mu} - \slashed{q}p^{\mu}_\Lambda)\bigr] p^{\mu_1}_\Lambda p^{\mu_2}_\Lambda p^{\mu_3}_\Lambda p^{\mu_4}_\Lambda 
\end{eqnarray}
and 
\begin{eqnarray}
  \label{eq:em_11}
\Gamma_{\rm{em}}^{\nu\nu_1\nu_2\nu_3\nu_4\pm} &=& \frac{-i}{m_{R}^6} \biggl[g_1p^{\nu}(\slashed{k}\slashed{\epsilon} -  \slashed{\epsilon}\slashed{k})+ g_2(k^{\nu}p\cdot \epsilon - \epsilon^{\nu}p\cdot k)+g_3 (\epsilon^\nu\slashed{k} - k^\nu\slashed{\epsilon})\slashed{p} +g_4 \gamma^{\nu} (\slashed{k}\slashed{\epsilon} -  \slashed{\epsilon}\slashed{k})\slashed{p}  \nonumber \\ & &  + g_5\gamma^{\nu}(p\cdot\epsilon\slashed{k} - p\cdot k\slashed{\epsilon})\biggr] k^{\nu_1}k^{\nu_2}k^{\nu_3}k^{\nu_4}\Gamma_{\pm}~ ,
\end{eqnarray}
respectively, whereas for the spin-13/2 resonance
\begin{eqnarray}
  \label{eq:had_13}
\Gamma_{\rm{had}}^{\mu\mu_1\mu_2\mu_3\mu_4\mu_5\pm}&=& \frac{g_{KYR}}{m_{R}^7}\Gamma_{\pm}\bigl[(p_\Lambda\cdot q - \slashed{p}_\Lambda\slashed{q})\gamma^\mu + \slashed{p}_\Lambda q^{\mu} - \slashed{q}p^{\mu}_\Lambda)\bigr] p^{\mu_1}_\Lambda p^{\mu_2}_\Lambda p^{\mu_3}_\Lambda p^{\mu_4}_\Lambda p^{\mu_5}_\Lambda ~,
\end{eqnarray}
and 
\begin{eqnarray}
  \label{eq:em_13}
\Gamma_{\rm{em}}^{\nu\nu_1\nu_2\nu_3\nu_4\nu_5\pm} &=& \frac{-i}{m_{R}^7} \biggl[g_1p^{\nu}(\slashed{k}\slashed{\epsilon} -  \slashed{\epsilon}\slashed{k})+ g_2(k^{\nu}p\cdot \epsilon - \epsilon^{\nu}p\cdot k)+g_3 (\epsilon^\nu\slashed{k} - k^\nu\slashed{\epsilon})\slashed{p} +g_4 \gamma^{\nu} (\slashed{k}\slashed{\epsilon} -  \slashed{\epsilon}\slashed{k})\slashed{p} \nonumber \\ 
& & + g_5\gamma^{\nu}(p\cdot\epsilon\slashed{k} - p\cdot k\slashed{\epsilon})\biggr]k^{\nu_1}k^{\nu_2}k^{\nu_3}k^{\nu_4}k^{\nu_5}\Gamma_{\mp}~, 
\end{eqnarray}
respectively. The above result indicates that the number of momentum dependence 
increases with the number of spin. This conclusion was previously made by Ref.~\cite{Huang:2004we}.

For the spin-15/2 resonance the hadronic and electromagnetic vertices are
given by 
\begin{eqnarray}
  \label{eq:had_15}
\Gamma_{\rm{had}}^{\mu\mu_1\mu_2\mu_3\mu_4\mu_5\mu_6\pm}&=& \frac{g_{KYR}}{m_{R}^8}\Gamma_{\pm}\bigl[(p_\Lambda\cdot q - \slashed{p}_\Lambda\slashed{q})\gamma^\mu + \slashed{p}_\Lambda q^{\mu} - \slashed{q}p^{\mu}_\Lambda)\bigr]
p^{\mu_1}_\Lambda p^{\mu_2}_\Lambda p^{\mu_3}_\Lambda p^{\mu_4}_\Lambda p^{\mu_5}_\Lambda p^{\mu_6}_\Lambda ~,
\end{eqnarray}
and
\begin{eqnarray}
  \label{eq:em_15}
\Gamma_{\rm{em}}^{\nu\nu_1\nu_2\nu_3\nu_4\nu_5\nu_6\pm} &=& \frac{-i}{m_{R}^8} \biggl[g_1p^{\nu}(\slashed{k}\slashed{\epsilon} -  \slashed{\epsilon}\slashed{k})+ g_2(k^{\nu}p\cdot \epsilon - \epsilon^{\nu}p\cdot k)
+g_3 (\epsilon^\nu\slashed{k} - k^\nu\slashed{\epsilon})\slashed{p}  +g_4 \gamma^{\nu} (\slashed{k}\slashed{\epsilon} -  \slashed{\epsilon}\slashed{k})\slashed{p} 
\nonumber\\
&& + g_5\gamma^{\nu}(p\cdot\epsilon\slashed{k} - p\cdot k\slashed{\epsilon})\biggr]
k^{\nu_1}k^{\nu_2}k^{\nu_3}k^{\nu_4}k^{\nu_5}k^{\nu_6}\Gamma_{\mp}~,
\end{eqnarray}
respectively. 
The above formalism is valid for both positive and negative parities, for which 
the parity factors are denoted by $\Gamma_{+} = i\gamma_5$ and $\Gamma_{-} = 1$, respectively.

\subsection{Production Amplitudes}
As stated above the production amplitudes for spin-11/2, -13/2, and -15/2 resonances
are obtained by sandwiching the propagators 
given by Eqs.~(\ref{eq:prop_11}), (\ref{eq:prop_13}), and (\ref{eq:prop_15}) 
between the corresponding hadronic vertex factors given by Eqs.~(\ref{eq:had_11}), 
(\ref{eq:had_13}), and (\ref{eq:had_15}), 
and electromagnetic vertex factors given by Eqs.~(\ref{eq:em_11}), (\ref{eq:em_13}), 
and (\ref{eq:em_15}), respectively.
As a result we obtain the amplitude for spin-11/2
\begin{eqnarray}
\label{eq:m112}
\mathcal{M}^{\pm}_{11/2} &=& \bar{u}_\Lambda \gamma_5 \{-s \pm m_{R}(\slashed{p}+\slashed{k})\}
\Bigl[7(33c_1^4+18c_1^2c_2c_3+7c_2^2c_3^2) \times\bigl\{-p_{\Lambda\nu}+{\frac{1}{s}}
c_\Lambda (p+k)_\nu\bigr\} -14c_1c_2(6c_1^2 + 2c_2c_3)
\nonumber \\
& & \times\bigl\{-k_\nu +{\frac{1}{s}}c_k (p+k)_\nu\bigr\}
+(21c_1^4 + 14c_1^2c_2c_3+c_2^2c_3^2)\bigl\{-\slashed{p}_\Lambda +{\frac{1}{s}}c_\Lambda(\slashed{p}+\slashed{k})\bigr\}\bigl\{-\gamma _\nu +{\frac{1}{s}}(\slashed{p}+\slashed{k}) (p+k)_\nu\bigr\}\nonumber \\
& &-14c_1(6c_1^2 + 2c_2c_3) \bigl\{-\slashed{p}_\Lambda +{\frac{1}{s}}c_\Lambda(\slashed{p}+\slashed{k})\bigr\}\bigl\{-\slashed{k}+{\frac{1}{s}}c_k(\slashed{p}+\slashed{k})\bigr\}
\bigl\{-p_{\Lambda\nu}+{\frac{1}{s}}c_\Lambda (p+k)_\nu\bigr\}+ 4c_2(7c_1^2 + c_2c_3) \nonumber \\
& & \times\bigl\{-\slashed{p}_\Lambda +{\frac{1}{s}}c_\Lambda(\slashed{p}+\slashed{k})\bigr\}
\bigl\{-\slashed{k}+{\frac{1}{s}}c_k(\slashed{p}+\slashed{k})\bigr\} \bigl\{-k_\nu +{\frac{1}{s}}c_k (p+k)_\nu\bigr\}\Bigr]\nonumber\\
& &\times \Bigl[ G_{1a} p^\nu (\slashed{k}\slashed{\epsilon}-\slashed{\epsilon}\slashed{k}) + G_{2a} (k^\nu p\cdot\epsilon - \epsilon^\nu p\cdot k)+ G_{3a} (\epsilon^\nu \slashed{k} -k^\nu \slashed{\epsilon} )\slashed{p}\Bigr]u_p \, ,
\end{eqnarray}
with: 
\begin{eqnarray}
G_{1a} &=& \frac{s^5g_{KYR} g_1}{231m^{22}_{R}(s-m^2_{R}+im_{R}\Gamma_{R})} \, ,\\
G_{2a} &=& \frac{s^5g_{KYR} g_2}{231m^{22}_{R}(s-m^2_{R}+im_{R}\Gamma_{R})} \, ,\\
G_{3a} &=& \frac{s^5g_{KYR} g_3}{231m^{22}_{R}(s-m^2_{R}+im_{R}\Gamma_{R})} \, .
\end{eqnarray} 

The production amplitude for spin-13/2 particle is given by 
\begin{eqnarray}
\label{eq:m132}
\mathcal{M}^{\pm}_{13/2} &=& \bar{u}_\Lambda \gamma_5 \{s \pm m_{R}(\slashed{p}+\slashed{k})\}\Bigl[3c_1(143c_1^4 + 110c_1^2c_2c_3+15c_2^2c_3^2) \bigl\{-p_{\Lambda\nu}+{\frac{1}{s}}c_\Lambda (p+k)_\nu\bigr\}-5c_2(33c_1^4
\nonumber\\ & & +18c_1^2c_2c_3+c_2^2c_3^2) \bigl\{-k_\nu +{\frac{1}{s}}c_k (p+k)_\nu\bigr\} +c_1(33c_1^4+120c_1^2c_2c_3+5c_2^2c_3^2) \bigl\{-\slashed{p}_\Lambda +{\frac{1}{s}}c_\Lambda(\slashed{p}+\slashed{k})\bigr\}
\nonumber\\ & & \times \bigl\{-\gamma _\nu +{\frac{1}{s}}(\slashed{p}+\slashed{k}) (p+k)_\nu\bigr\}
 -5(33c_1^4+18c_1^2c_2c_3+c_2^2c_3^2)\bigl\{-\slashed{p}_\Lambda +{\frac{1}{s}}c_\Lambda(\slashed{p}+\slashed{k})\bigr\}\bigl\{-\slashed{k}+{\frac{1}{s}}c_k(\slashed{p}+\slashed{k})\bigr\}
\nonumber\\ & &
\times\bigl\{-p_{\Lambda\nu}+{\frac{1}{s}}c_\Lambda (p+k)_\nu\bigr\}+20c_1c_2(3c_1^2 + c_2c_3)\bigl\{-\slashed{p}_\Lambda +{\frac{1}{s}}c_\Lambda(\slashed{p}+\slashed{k})\bigr\}
\bigl\{-\slashed{k}+{\frac{1}{s}}c_k(\slashed{p}+\slashed{k})\bigr\} 
\nonumber\\ & &
\times\bigl\{-k_\nu +{\frac{1}{s}}c_k (p+k)_\nu\bigr\}\Bigr]
 \Bigl[ G_{1b} p^\nu (\slashed{k}\slashed{\epsilon}-\slashed{\epsilon}\slashed{k}) + G_{2b} (k^\nu p\cdot\epsilon - \epsilon^\nu p\cdot k)+ G_{3b} (\epsilon^\nu \slashed{k} -k^\nu \slashed{\epsilon} )\slashed{p}\Bigr]u_p \, ,
\end{eqnarray}
with
\begin{eqnarray}
G_{1b} &=& \frac{s^6g_{KYR} g_1}{429m^{26}_{R}(s-m^2_{R}+im_{R}\Gamma_{R})}\, ,\\
G_{2b} &=& \frac{s^6g_{KYR} g_2}{429m^{26}_{R}(s-m^2_{R}+im_{R}\Gamma_{R})}\, ,\\
G_{3b} &=& \frac{s^6g_{KYR} g_3}{429m^{26}_{R}(s-m^2_{R}+im_{R}\Gamma_{R})}\, .
\end{eqnarray}
Finally, the amplitude for spin-15/2 resonance reads 
\begin{eqnarray}
\label{eq:m152}
\mathcal{M}^{\pm}_{15/2} &=& \bar{u}_\Lambda \gamma_5 \{-s \pm m_{R}(\slashed{p}+\slashed{k})\}\Bigl[45(143c_1^6+143c_1^4c_2c_3+33c_1^2c_2^2c_3^2 + c_2^3c_3^3)
\bigl\{-p_{\Lambda\nu}+{\frac{1}{s}}c_\Lambda (p+k)_\nu\bigr\} \nonumber \\
& & -9c_1c_2(286c_1^4 + 220c_1^2c_2c_3+30c_2^2c_3^2)
\bigl\{-k_\nu +{\frac{1}{s}}c_k (p+k)_\nu\bigr\} +(429c_1^6 + 495c_1^4c_2c_3 + 135c_1^2c_2^2c_3^2)\nonumber \\
& & \times \bigl\{-\slashed{p}_\Lambda +{\frac{1}{s}}c_\Lambda(\slashed{p}+\slashed{k})\bigr\}\bigl\{-\gamma _\nu +{\frac{1}{s}}(\slashed{p}+\slashed{k}) (p+k)_\nu\bigr\}
-18c_1(143c_1^4 + 110c_1c_2c_3 + 15c_2^2c_3^2) \nonumber \\
& & \times\bigl\{-\slashed{p}_\Lambda +{\frac{1}{s}}c_\Lambda(\slashed{p}+\slashed{k})\bigr\}\bigl\{-\slashed{k}+{\frac{1}{s}}c_k(\slashed{p}+\slashed{k})\bigr\}\bigl\{-p_{\Lambda\nu}+{\frac{1}{s}}c_\Lambda (p+k)_\nu\bigr\}
 + 30c_2(33c_1^4 +18c_1c_2c_3^2 + c_2^2c_3^2)\nonumber\\ 
& & \times\bigl\{-\slashed{p}_\Lambda +{\frac{1}{s}}c_\Lambda(\slashed{p}+\slashed{k})\bigr\}
\bigl\{-\slashed{k}+{\frac{1}{s}}c_k(\slashed{p}+\slashed{k})\bigr\}\bigl\{-k_\nu +{\frac{1}{s}}c_k (p+k)_\nu\bigr\}\Bigr]
\nonumber\\ 
& & \times\Bigl[ G_{1c} p^\nu (\slashed{k}\slashed{\epsilon}-\slashed{\epsilon}\slashed{k}) + G_{2c} (k^\nu p\cdot\epsilon - \epsilon^\nu p\cdot k)
+ G_{3c} (\epsilon^\nu \slashed{k} -k^\nu \slashed{\epsilon} )\slashed{p}\Bigr]u_p \, ,
\end{eqnarray}
\end{widetext}
with
\begin{eqnarray}
G_{1c} &=& \frac{s^7g_{KYR} g_1}{6435m^{30}_{R}(s-m^2_{R}+im_{R}\Gamma_{R})}\, ,\\
G_{2c} &=& \frac{s^7g_{KYR} g_2}{6435m^{30}_{R}(s-m^2_{R}+im_{R}\Gamma_{R})}\, ,\\
G_{3c} &=& \frac{s^7g_{KYR} g_3}{6435m^{30}_{R}(s-m^2_{R}+im_{R}\Gamma_{R})}\, .
\end{eqnarray}
Note that in the fitting process only the product of the hadronic and electromagnetic couplings, i.e.,
$g_{KYR} g_i$ with $i=1,2,3$, are extracted from the data. Furthermore, in the production amplitudes
given by Eqs.~(\ref{eq:m112}), (\ref{eq:m132}), and (\ref{eq:m152}) we have used the following definitions,
\begin{subequations}
\label{eq:c_defined}
\begin{eqnarray}
	c_1  =  b_{\Lambda} - c_{\Lambda}c_k/s \, , \\
	c_2  =  m_{\Lambda}^2 - c_{\Lambda}^2/s \, , \\
	c_3  =  c_k^2/s - k^2 \, , \\
	c_4  =  2b_p + k^2 \, ,\\
	c_5  =  4b_p + k^2 \, ,\\
	b_p  =  p \cdot k \, ,\\
	b_{\Lambda}  =  p_{\Lambda} \cdot k \, ,\\
	b_q  =  q \cdot k \, ,\\
	c_p  =  (p + k) \cdot p \, ,\\
	c_{\Lambda}  =  (p + k) \cdot p_{\Lambda} \, ,\\
        c_k  =  (p + k) \cdot k \,, \\
	c_s  =  1 - c_{\Lambda}/s\, .
\end{eqnarray}
\end{subequations}

\subsection{Calculation of the observables}
The production amplitudes given by Eqs.~(\ref{eq:m112}), (\ref{eq:m132}), and (\ref{eq:m152})
can be decomposed into six gauge and Lorentz invariant matrices $M_i$ through 
\begin{eqnarray}\label{inv}
\mathcal{M}_{fi}=\bar{u}_\Lambda\sum^{6}_{i=1} A_i M_i\, u_p ~,
\end{eqnarray}
where the gauge and Lorentz invariant matrices $M_i$ are given by~\cite{Deo:1974ik,Dennery:1961zz}:
\begin{eqnarray}
M_1&=& \gamma_5 \,\slashed{\epsilon}\slashed{k}  ~,\\
M_2&=& 2\gamma_5\left( q\cdot\epsilon P\cdot k - q\cdot k\, P\cdot\epsilon\right) ~,\\
M_3&=& \gamma_5(q\cdot k\slashed{\epsilon}-q\cdot\epsilon \slashed{k}) ~,\\
M_4&=& i\varepsilon_{\mu\nu\rho\sigma}\gamma^\mu q^\nu\epsilon ^\rho k^\sigma ~,\\
M_5&=& \gamma_5(q\cdot\epsilon k^2 - q\cdot k k \cdot\epsilon), \\
M_6&=& \gamma_5(k \cdot \epsilon \slashed{k} - k^2 \slashed{\epsilon}) ~,
\end{eqnarray} 
where $P = \frac{1}{2}(p + p_\Lambda)$ and  $\varepsilon_{\mu\nu\rho\sigma}$ is 
the Levi-Civita antisymmetric tensor. All observables required for fitting the
experimental data can be calculated from the form function $A_i$ extracted from 
Eq.~(\ref{inv}), after adding the contributions from all involved intermediate states.
The form functions $A_i$ for baryon resonances with spins up to 9/2 are given
in the previous works \cite{Clymton:2017nvp,Mart:2015jof}, whereas those with
spins 11/2, 13/2, and 15/2 considered in the present work 
are given in Appendix~\ref{app:form_function}.

\subsection{Pole Position}
Besides the Breit-Wigner parameters, such as mass, width, and branching ratios, 
the Particle Data Group has recently listed new information on the resonance properties, 
i.e., the pole position. It is obvious that the Breit-Wigner parameters extracted in 
each model depend highly on the background terms of the model. Therefore, all resonance 
properties obtained by using such parameterization is difficult to compare with those
obtained from other models. This problem does not appear in the case of pole position. 
Currently, the pole position has been extensively used in the realm of hadronic physics. 
In the Particle Data Book 2018 the pole positions of resonance are 
listed before the Breit-Wigner parameters. The placement shows that the pole positions
is currently considered as the important properties of a resonance. 

In principle, the pole position can be calculated by setting the denominator of the 
scattering amplitude to zero. Approaching the pole position the scattering amplitude 
of a resonance increases dramatically. Since the resonance scattering amplitude 
becomes extremely larger than contributions from other intermediate states, 
the resonance property calculated at the pole position is insensitive to the 
contribution of  background terms. As a result, the evaluation of  resonance 
properties at the pole position is practically model independent. 

In the present work, the pole position properties of a resonance are the resonance
mass and width. They are defined {\it via}
\begin{equation}
\sqrt{s_R} = M_{\rm pole} - i\Gamma_{\rm pole}/2 \, .
\end{equation}
As previously stated, this is obtained by setting the denominator of
scattering amplitude to zero, i.e.,
\begin{equation}
\label{eq:pole_set_to_zero}
s_R - m_R^2 + im_R \Gamma(s_R) = 0 \, .
\end{equation}
Notice that the above equation cannot be directly calculated, since in the present study 
we use $\Gamma(s)$ that depends on the total c.m. energy. Therefore, the solution of 
Eq.~(\ref{eq:pole_set_to_zero}) must be obtained numerically.

\section{Results and Discussion}
\label{sec:result}
In the present work, the isobar model used to analyze the effect of spin-11/2 and -13/2 nucleon 
resonances in the $K\Lambda$ channels is based on our previous model developed to describe 
all available data in these channels~\cite{Mart:2019mtq}. Furthermore, in this study we also 
investigate the effect of spin-11/2, -13/2, and -15/2 $\Delta$ resonances in the $K\Sigma$ 
reaction channels. Along with the nucleon resonances used in the $K\Lambda$ channels, these
$\Delta$ resonances are listed in Table~\ref{tab:resonance}. In total, there are 23 nucleon 
resonances included in our analysis for the $K\Lambda$ and $K\Sigma$ channels and, in addition, 
17 $\Delta$ resonances in the $K\Sigma$ channels with spins up to spin-15/2. The result obtained 
in all channels will be discussed in the following subsections.
\subsection{$K\Lambda$ Channel}
 \begin{table}[!h]
   \caption{Coupling constants and other driving parameters of the background terms
     for $K\Lambda$ channels obtained in the present work (Model A) and the previous
     one (Model B) \cite{Mart:2019mtq}. Error bars were not reported in Model B.
     See Ref.~\cite{Mart:2019mtq} for the explanation
     of the parameter notation.}
   \label{tab:par1}
   \begin{ruledtabular}
     \begin{tabular}[c]{lrr}
       Parameter & Model A & Model B  \\
       \hline
       $g_{K \Lambda N} / \sqrt{4 \pi}$ & $-4.40\pm 0.03$  & $ -3.00$ \\
       $g_{K \Sigma N} / \sqrt{4 \pi}$  & $0.90 \pm 0.04$  & $  1.30$ \\
       $G^{V}_{K^*} / 4\pi$             & $0.08 \pm 0.00$  & $  0.13$ \\
       $G^{T}_{K^*} / 4\pi$             & $-0.07\pm 0.00$  & $  0.17$ \\
       $G^{V}_{K_1} / 4\pi$             & $0.12 \pm 0.00$  & $  0.13$ \\
       $G^{T}_{K_1} / 4\pi$             & $2.43 \pm 0.00$  & $  3.89$ \\
       $r_{{K_1}{K_\gamma}}$            & $0.52 \pm 0.01$  & $  0.65$ \\
       $\Lambda_{\rm B}$ (GeV)          & $0.89 \pm 0.00$  & $  0.70$ \\
       $\Lambda_{\rm R}$ (GeV)          & $1.09 \pm 0.00$  & $  1.10$ \\
       $\theta_{\rm had}$ (deg)         & $94.31 \pm 0.64$  & $  90.0$ \\
       $\phi_{\rm had}$ (deg)           & $90.00 \pm 4.09$  & $  0.0 $ \\
       \hline 
       $\chi^2$                         & 13316   & 13867   \\
       $N_{\rm par}$                    & 264     & 247     \\
       $N_{\rm data}$                   & 9364    & 9364    \\
       $\chi^2 / N_{\rm dof}$                     & 1.46    & 1.52    \\
     \end{tabular}
   \end{ruledtabular}
 \end{table}

In Table~\ref{tab:par1} we present the leading coupling constants and other background parameters 
obtained from the previous work (Model B) \cite{Mart:2019mtq} and current analysis (Model A). 
Note that in the present work we have omitted the $K^0\Lambda$ photoproduction data obtained 
from MAMI collaboration~\cite{mami2018} due to the problem of data discrepancy as discussed
in Ref.~\cite{Mart:2019mtq}. Furthermore, it was shown that by excluding these data from 
the database leads to a better model that can nicely reproduce the $\gamma n\to K^0\Lambda$ 
helicity asymmetry $E$ \cite{Mart:2019mtq}. It is important to note that Model B was also 
obtained from fitting without these data. 

From Table~\ref{tab:par1} we can conclude that there is no dramatic changes in the background
parameters after including the spin-11/2 and -13/2 nucleon resonances in the model. Nevertheless,
the increase of $g_{K \Lambda N}$ coupling and the Born hadronic cutoff $\Lambda_{\rm B}$
shows that the inclusion of the two resonances increases the contribution of the background
terms. We note that in the case of Kaon-Maid, the Born cutoff is very soft, i.e., 
$\Lambda_{\rm B}=0.637$ GeV \cite{kaon-maid}. Clearly, the Kaon-Maid model is  
dominated by the resonance terms, whereas the Born terms are strongly suppressed. 
Such situation is completely different from the case of pion or eta photoproduction
and could raise a question, whether Kaon-Maid is a realistic phenomenological model.

The listed $\chi^2$ values indicate that the agreement 
between model calculation and experimental data is significantly improved after including 
the two nucleon resonances, as clearly expected.
Since the calculation includes two isospin channels, i.e., $\gamma p \to K^+ \Lambda$ and 
$\gamma n \to K^0 \Lambda$, in the followings we present comparison between model calculations 
and experimental data in details. 

\subsubsection{$K^+\Lambda$ Channel}
\begin{figure}[!t]
   \centering
   \includegraphics[scale=0.45]{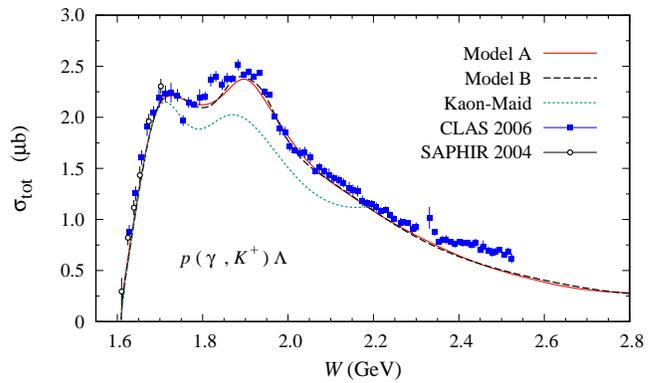}
   \caption{Calculated total cross section of the $\gamma p \rightarrow K^+ \Lambda$ channel obtained from previous~\cite{Mart:2019mtq} and present works, compared with the experimental data from the CLAS collaboration (solid squares \cite{clas2006}).}
   \label{fig:total1}
 \end{figure}

Comparison between calculated $\gamma p \to K^+ \Lambda$ total cross sections 
from Models A and B, Kaon-Maid, and experimental data is displayed in 
Fig.~\ref{fig:total1}. Note that the experimental data shown in this figure 
are only for visual comparison. The data were not included in the fitting process, 
since differential and total cross sections data come from the same experiment. 

\begin{figure}[!b]
	\centering
	\includegraphics[scale=0.7]{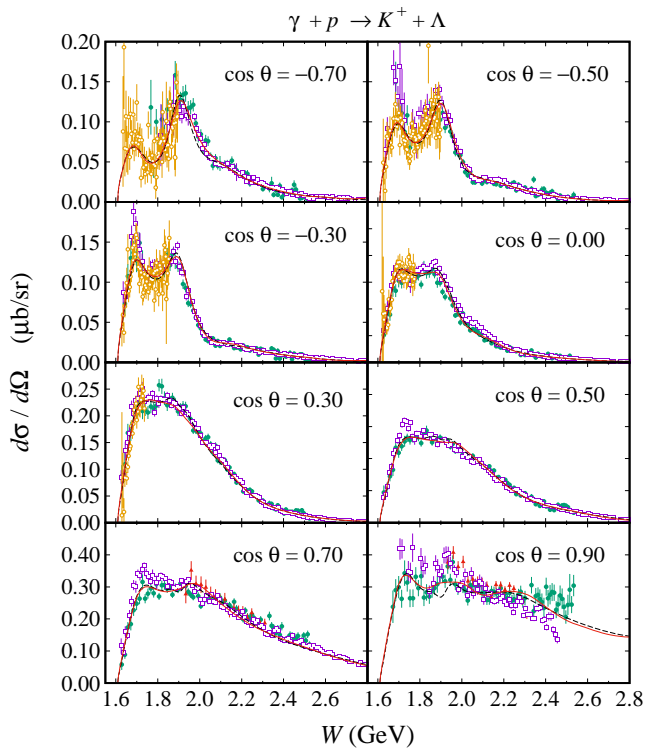}
	\caption{Energy distributions of the $\gamma p \rightarrow K^+ \Lambda$ 
          differential cross section obtained from Model A (solid red curves) 
          and Model B (dashed black curves) for different values of $\cos\theta$. 
          Experimental data shown in this figure are
          obtained from the LEPS 2006 (solid triangles \cite{leps2006}), 
          CLAS 2006 (solid squares \cite{clas2006}), 
          CLAS 2010 (solid circles \cite{clas2010}), and 
          Crystal Ball 2014 (open circles \cite{crystalball})
          collaborations. }
	\label{fig:dkpl_e}
\end{figure}

\begin{figure}[!]
	\centering
	\includegraphics[scale=0.7]{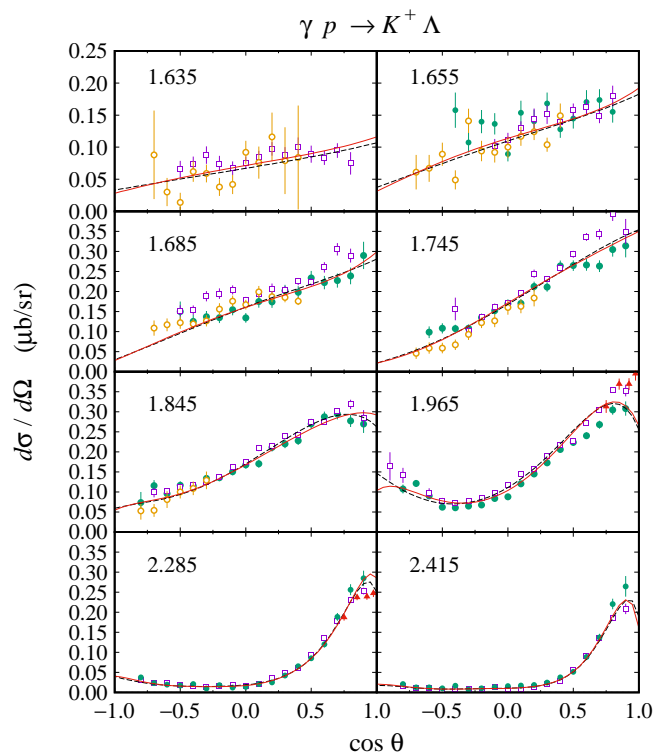}
	\caption{As in Fig.~\ref{fig:dkpl_e}, but for angular distribution.
        The corresponding value of total c.m. energy $W$ in GeV is shown in each panel.}
	\label{fig:dkpl_th}
\end{figure}

Compared to the prediction of Kaon-Maid, both models A and B displayed in Fig.~\ref{fig:total1}
show substantial improvement. However, since our main motivation in this work
is to investigate the effect of spin-11/2 and 13/2 nucleon resonance, we will 
not compare our result with the prediction of Kaon-Maid in the following discussion,
except in the case of total cross section, in which recent experimental data are
in good agreement with Kaon-Maid for certain isospin channel. Both current and previous models 
seem to have a great agreement, with a tiny difference 
only at higher energy region, i.e., $W \geq 2.6$ GeV. The difference originates from
the use of the high spin nucleon resonances, as obviously seen from their masses.
However, since there are no available 
data in this energy region, no conclusion can be drawn at this point. Future experiments with 
12 GeV electron source at JLab could be expected to reveal more information in this
energy regime.

Both peaks shown by 
the two models seem to agree with each other, with minuscule difference at $W \approx 1.85$ 
GeV, where the second peak is attributed to the $P_{13}(1990)$ state, as discussed in 
Ref.~\cite{Mart:2012fa}. 

More information can be obtained from the differential cross sections shown in 
Figs.~\ref{fig:dkpl_e} and \ref{fig:dkpl_th}. Figure~\ref{fig:dkpl_e} shows that
the difference in the total cross sections of models A and B originate from the
forward and backward regions of the differential cross section. Furthermore, 
as shown in Fig.~\ref{fig:dkpl_e} at backward angle (cos $\theta = -0.70$), 
it is also apparent that the result of 
model A has a better agreement than model B, except in the higher energy region,
$W>2.4$ GeV. Another interesting result is that the second peak of Model A is 
slightly shorter, but wider, than that of Model B. As a result, Model A yields
a more accurate explanation of experimental data, especially for the CLAS 
2010~\cite{clas2010} and Crystal Ball~\cite{crystalball} ones. However, 
in the forward region Model A yields fewer peaks than Model B. Nevertheless,
Model A seems to produce more natural shape of the cross section at the very
forward angle, $\cos\theta=0.90$, where unfortunately, the available experimental data from
different collaborations produce uncertainty in differential 
cross section up to nearly 40\%.

The angular distribution of differential cross section displayed in Fig.~\ref{fig:dkpl_th} 
shows that both models are in good agreement with experimental data. Furthermore, experimental 
data in higher energy region are better reproduced. This figure again shows that the 
inclusion of the high-spin resonances with higher masses does not influence the
the cross section behavior in the lower energy region, where experimental data exist. 

\begin{figure*}[!]
  \includegraphics[scale=0.6]{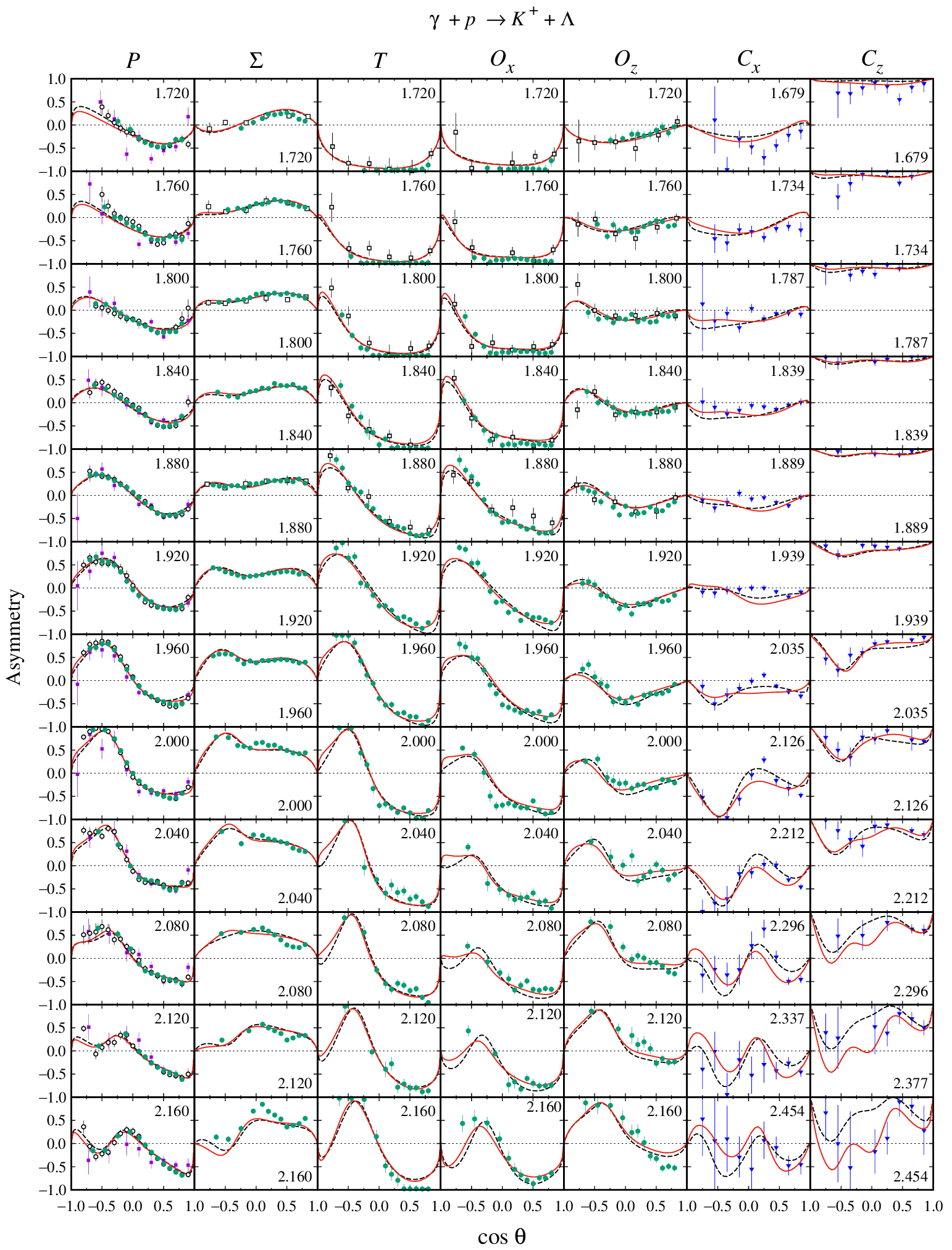}
\caption{Asymmetry of the single-polarization ($P$, $\Sigma$, and $T$), and 
  double-polarization ($O_x$,  $O_z$,  $C_x$ and  $C_z$) observables 
  for the $\gamma p \to K^+\Lambda$ channel as a function of kaon angle
  for different total c.m. energies $W$ shown in each panel.  Dashed black
  curves are obtained from the previous isobar model \cite{Mart:2019mtq}, 
  solid red curves are obtained from the present calculation. 
  Experimental data shown in this figure are taken from 
  the GRAAL 2009 (open squares \cite{graal2009}), 
  CLAS 2006 (solid squares \cite{clas2006}),
  CLAS 2007 (solid triangles \cite{clas2007}), 
  CLAS 2010 (open circles \cite{clas2010}), and 
  CLAS 2016 (solid circles \cite{clas2016}) collaborations.}
\label{fig:clas_new_pol}
\end{figure*}

Figure~\ref{fig:clas_new_pol} shows the single- and double-polarization observables,
for which experimental data are abundantly available at present. In this case we
do not see a dramatic changes after including the spin-11/2 and 13/2 nucleon resonance
in the model, except improvement in the agreement between model calculation and 
experimental data at forward
direction and high energy region, which is expected due to the higher masses of these
resonances. Nevertheless, we still see significant improvement in the beam-recoil
double-polarization observables $C_x$ and $C_z$ at higher energies, where experimental
data have large error bars. Note that only a small part of experimental data can be 
displayed in Fig.~\ref{fig:clas_new_pol}. More data are available in the fitting database,
especially for the recoil polarization $P$, and are not shown in the figure due to their
different kinematics. 

\subsubsection{$K^0\Lambda$ Channel}

The available data for the $K^0\Lambda$ channel are significantly fewer than the $K^+\Lambda$ 
one, given that the experiment with neutron target is more difficult to perform. The number 
of data included in this study is less than 1000, which will affect the accuracy between 
model calculation and experimental data. The calculation of this channel is performed by 
using the isospin relation of the hadronic coupling constants, as well as information on the
neutral kaon transition moment and neutron helicity photon coupling obtained from PDG
\cite{pdg}, in the $K^+ \Lambda$ model as discussed in 
Refs.~\cite{Mart:1995wu,Mart_K0L:2011}. 
Thus, investigation of the $K^0\Lambda$ channel can be considered as a direct test 
of isospin symmetry in kaon photoproduction.

Experimental data of this channel are already available from the CLAS g10 and g13 
collaboration~\cite{clas2017} and MAMI 2018 collaboration~\cite{mami2018}. 
In the previous work~\cite{Mart:2019mtq}, both data sets were included in the 
analysis. However, in the present work we exclude the data from the MAMI 2018 
collaboration, since it was found that the data are more difficult to fit and
have a discrepancy problem with the CLAS g10 and g13 data. Furthermore, in the
present work our main motivation is to investigate the effect of the higher
spin nucleon resonances, in which we need an accurate isobar model.

\begin{figure}[h!]
  \centering
  \includegraphics[scale=0.45]{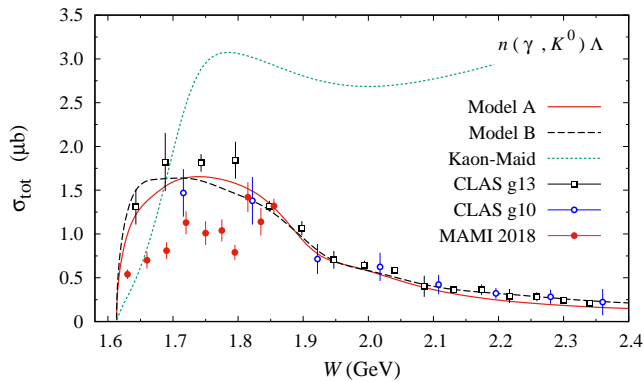}
  \caption{Total cross section of the $\gamma n \rightarrow K^0 \Lambda$
    isospin channel calculated from different models. Experimental data 
    are obtained from the CLAS g10 and g13 
    collaboration (open circles and open squares) \cite{clas2017} and 
    MAMI 2018 collaboration (solid circles) \cite{mami2018}. Note that
    the data shown in this figure were not used in the fitting process 
    and shown here only for comparison.}
  \label{fig:k0ltotal}
\end{figure}

Figure~\ref{fig:k0ltotal} shows the calculated total cross section of the  
$\gamma n \rightarrow K^0 \Lambda$ channel. Obviously, both models A and B 
yield similar cross section trend, except in the lower energy region and
especially near the production threshold, where the cross section obtained 
from model B is steeper than that of the present work. All models give the
total cross sections within the experimental error bars. As discussed in 
Ref.~\cite{Mart_K0L:2011}, threshold behavior of $K^0\Lambda$ photoproduction
provides important information that can shed more light on the difference
between pseudovector and pseudoscalar theories in the kaon photoproduction
process. The absence of $K^0$ exchange in this channel also reduces the 
number of unknown parameters in the model. As a consequence, threshold 
properties of the $K^0\Lambda$ can be more accurately investigated. Note
also that the over prediction of Kaon-Maid model is understandable, since it is
pure prediction and the model was fitted to old data.

\begin{figure}[h!]
  \centering
  \includegraphics[scale=0.57]{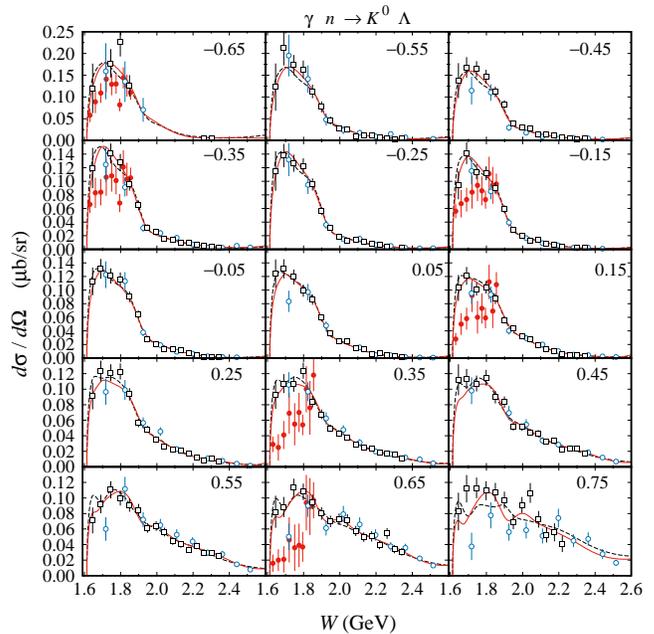}
  \caption{Energy distribution of the $\gamma n \rightarrow K^0 \Lambda$
    differential cross section for different values of $\cos\theta$ shown
    in each panel. Notation of the curves and experimental 
    data is as in Fig.~\ref{fig:k0ltotal}. }
  \label{fig:dk0l_e}
\end{figure}

\begin{figure}[t!]
	\centering
	\includegraphics[scale=0.57]{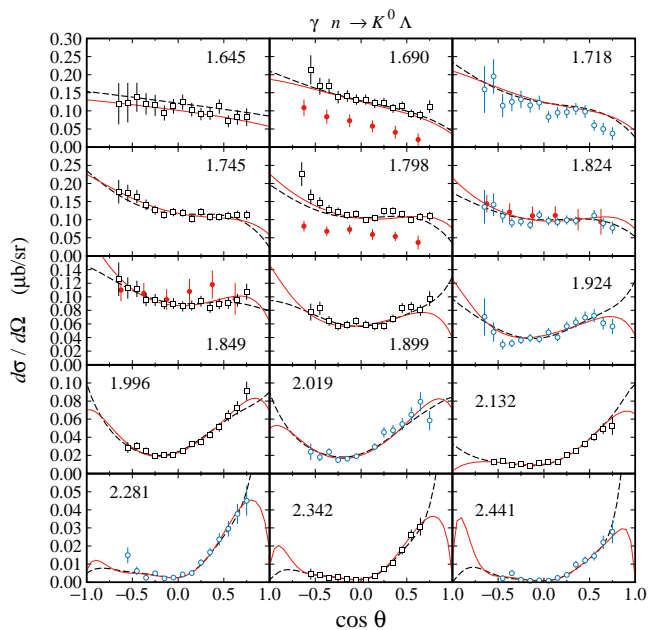}
	\caption{As in Fig.~\ref{fig:dk0l_e}, but for angular distributions.
        The total c.m. energy $W$ is shown in each panel in GeV.}
	\label{fig:dk0l_th}
\end{figure}

Figure~\ref{fig:dk0l_e} shows the differential cross section of the 
$\gamma n \rightarrow K^0 \Lambda$ channel. At a glance, both models 
seem to be similar, especially at $W > 1.8$ GeV region. However, the
inclusion of the two high-spin nucleon resonances leads to different
differential cross section near the threshold. The difference is more
apparent at the forward angle, i.e., $\cos\theta=0.75$. The angular
distribution of differential cross section shown in Fig.~\ref{fig:dk0l_th}
corroborates this result. In Fig.~\ref{fig:dk0l_th} we can see that the
difference between the two models is more obvious in the forward and
backward regions. 

In general, the $K^0\Lambda$ differential cross sections also show that the 
inclusion of the two high-spin nucleon resonances improves the model.
However, there is an important phenomenon appears in the $K^0\Lambda$ 
channel. As in the case of the $K^+\Lambda$ channel, the inclusion of
these resonances leads to fewer structures in differential cross section,
especially at the forward region (see Fig.~\ref{fig:dkpl_e}). The same
phenomenon is also displayed by the $K^0\Lambda$ channel as shown in
Fig.~\ref{fig:dk0l_e}. In this channel
the previous work displays a clear structure at $W\approx 1.65$ GeV, which
appears in a wide range of angular distribution, but is more apparent
near forward region. The structure is eliminated by the inclusion of the 
two high-spin nucleon resonances. We found that this structure
is very interesting because it originates from the $N(1650)$ resonance 
contribution and the corresponding width is less than 50 MeV (see the
dashed curves in Fig.~\ref{fig:dk0l_e}, especially at $\cos\theta=0.75$).
In the previous work \cite{Mart_K0L:2011} it was concluded that the structure
could be a hint of the narrow resonance, which was found to have the mass
of 1650 MeV.

\subsection{$K\Sigma$ Channel}
As in the $K\Lambda$ channels, the four available channels of $K\Sigma$ 
photoproduction can be also simultaneously analyzed by exploiting the 
isospin symmetry and some information on the resonance properties 
from PDG \cite{pdg}. Recently, we have studied 
these channels by using a partial wave approach for the resonance part, 
whereas the background part was still constructed from the covariant 
Feynman diagram technique \cite{Mart:2019fau}. The model was
fitted to nearly 8000 experimental data points available from all four 
channels, but dominantly from the $K^+\Sigma^0$ one. 

In addition, 
a fully covariant model to describe photoproduction of $K\Sigma$ 
has been also constructed by including nucleon resonances with spins 
up to 9/2 and the result has been submitted for publication
\cite{samson_ksigma}.
In the present work we add the nucleon and delta resonances that are not
available in this covariant model. They include the nucleon resonances 
with spins 11/2 and 13/2, as well as the delta resonances with spins 
11/2, 13/2, and 15/2, listed in Table \ref{tab:resonance}. The experimental 
data used in this study were obtained from the CLAS, Crystal Ball, GRAAL, 
SAPHIR, LEPS, and SPring8 collaborations. Thus, to observe the effect of
including these resonances, we will compare the result of our present 
work to that of the covariant model reported in Ref.~\cite{samson_ksigma}.

Table~\ref{tab:parametersig} lists the leading coupling constants and other 
background parameters extracted from the present analysis. For the sake of
discussion, the present model and the model reported in Ref.~\cite{samson_ksigma} 
will be referred to as Model C and Model D, respectively. As seen from the values 
of $\chi^2/N_{\rm dof}$ in Table~\ref{tab:parametersig}, the agreement between model 
calculation and experimental data is improved after the inclusion of the 
high-spin nucleon and delta resonances. The result is clearly expected,
because the addition of resonances increases the free parameters in the
model. Table~\ref{tab:parametersig} also shows that the inclusion of the 
high-spin resonances helps to increase the hadronic form factor cutoff of 
the Born terms as in the $K\Lambda$ case.

\begin{table}[!t]
  \centering
  \caption{Extracted coupling constants and other background parameters 
    in the $K\Sigma$ channels obtained from the present work (Model C) 
    and the previous one (Model D) \cite{samson_ksigma}.
    Note that error bars were not reported in Model D.}
  \label{tab:parametersig} 
  \begin{ruledtabular}
    \begin{tabular}[c]{lrr}
      Parameter & Model C & Model D\\
      \hline
      $g_{K \Lambda N} / \sqrt{4 \pi}$ & $-4.26 \pm 0.01 $  & $-3.00$  \\
      $g_{K \Sigma N} / \sqrt{4 \pi}$  & $1.30  \pm 0.25 $  & 0.90     \\
      $G^{V}_{K^*} / 4\pi$             & $-0.04 \pm 0.00 $  &  $-0.15$ \\
      $G^{T}_{K^*} / 4\pi$             & $-0.03 \pm 0.00 $  &  $-0.21$ \\
      $G^{V}_{K_1} / 4\pi$             & $-0.46 \pm 0.00 $  &    0.12   \\
      $G^{T}_{K_1} / 4\pi$             & $ 0.07 \pm 0.01 $  &    4.37  \\
      $r_{{K_1}{K_\gamma}}$            & $-2.00 \pm 0.36$  & $\cdots$\footnotemark[1] \\
      $\Lambda_{\rm B}$(GeV)           & $ 0.84 \pm 0.00 $  &  0.72    \\
      $\Lambda_{\rm R}$(GeV)           & $ 1.00 \pm 0.00 $  &  1.25    \\
      $\theta_{\rm had}$(deg)          & $ 90.00 \pm 10.11 $  & 90.0     \\
      $\phi_{\rm had}$(deg)            & $0.00 \pm 20.05 $  &   0.00   \\
      \hline
      $\chi^2$                         & 8729     & 9053     \\
      $N_{\rm par}$                    & 379      & 341      \\
      $N_{\rm data}$                   & 7784     & 7784     \\
      $\chi^2 / N_{\rm dof}$           & 1.18     & 1.22     \\
    \end{tabular}
  \footnotetext[1]{Not reported}
  \end{ruledtabular}
\end{table}

\subsubsection{$K^+\Sigma^0$ Channel}
Among all possible $K\Sigma$ isospin channels, the $\gamma p \to K^+ \Sigma^0$ channel 
has the most abundant experimental data. This is understandable since, as the 
production of $K^+\Lambda$, the production of $K^+\Sigma^0$ is relatively 
easier to measure due to stable proton target and relatively simpler technique 
to measure the decay of $\Lambda$ or $\Sigma^0$ hyperon in the final states. The 
experimental data mentioned here include those obtained from the Crystal Ball 
(at MAMI) \cite{crystalball}, CLAS \cite{clas2004sig,clas2006,biblap}, 
GRAAL \cite{graal:2007tx}, LEPS, SAPHIR and SPring8 collaborations. 

\begin{figure}[!t]
	\centering
	\includegraphics[scale=0.43]{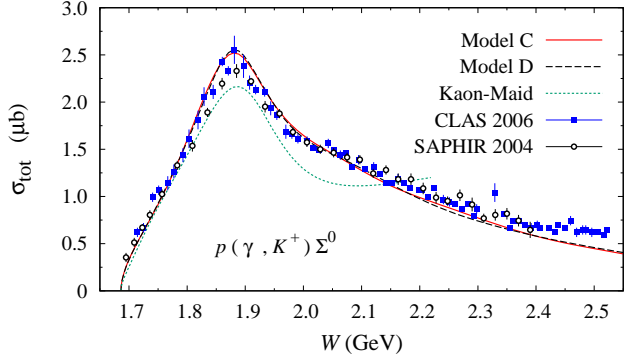}
	\caption{Total cross section of the $\gamma p \to K^+ \Sigma^0$ channel. 
        Notation of the curves and experimental is given in the figure. Data shown
        in this figure were not used in the fitting process of the present model.}
	\label{fig:total3}
\end{figure}

Figure~\ref{fig:total3} shows the comparison between the calculated total cross 
sections before and after the inclusion of the high-spin resonances, where the
prediction of Kaon-Maid is also displayed to show the improvement made by the 
current models. It is seen that the prediction of Model C (solid red curve) is 
practically similar to that of Model D (dashed black curve). The difference 
between both models is very subtle and can be seen only at $W \approx 2.0$ GeV 
and $W \approx 2.25$ GeV. Overall, both models fit nicely the experimental 
data, with an exception at $W \gtrsim 2.15$ GeV, where we observe that 
there is a discrepancy problem in the existing experimental data of total 
and differential cross sections. This problem will be clarified later when 
we discuss the result for differential cross section.

\begin{figure}[!b]
	\centering
	\includegraphics[scale=0.53]{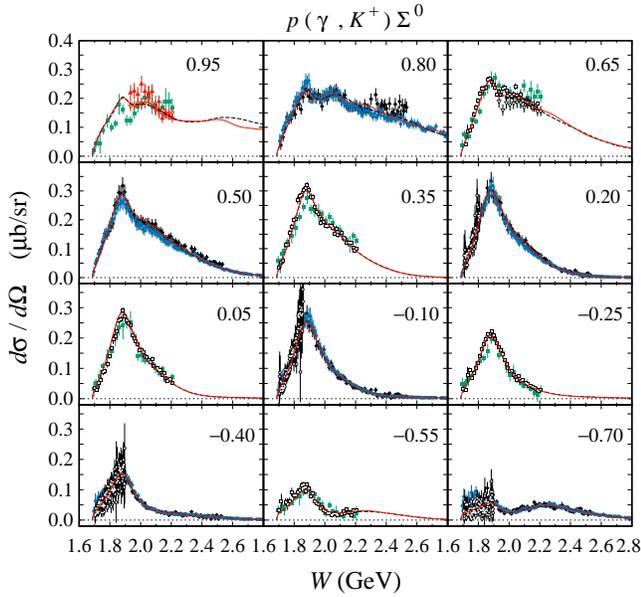}
	\caption{Energy distributions of the $\gamma p \to K^+ \Sigma^0$ 
          differential cross section for different values of $\cos\theta$. 
          Shown in the figure are the results 
          obtained from Model C (solid red curve) and Model D (dashed black 
          curve). Experimental data shown are taken from the CLAS 2004 
          (open squares \cite{clas2004sig}), SAPHIR 2004
          (solid squares \cite{sigsaphir2004}), CLAS 2006 (solid diamonds 
          \cite{clas2006}), LEPS 2006 (solid triangles \cite{leps2006} 
          and open inverted triangles \cite{sigleps2006}), CLAS 2007 
          (solid circles \cite{biblap}), and Crystal Ball 
          (open circles \cite{crystalball}) collaborations.}
	\label{fig:dkps_e}
\end{figure}

\begin{figure}[t!]
	\centering
	\includegraphics[scale=0.53]{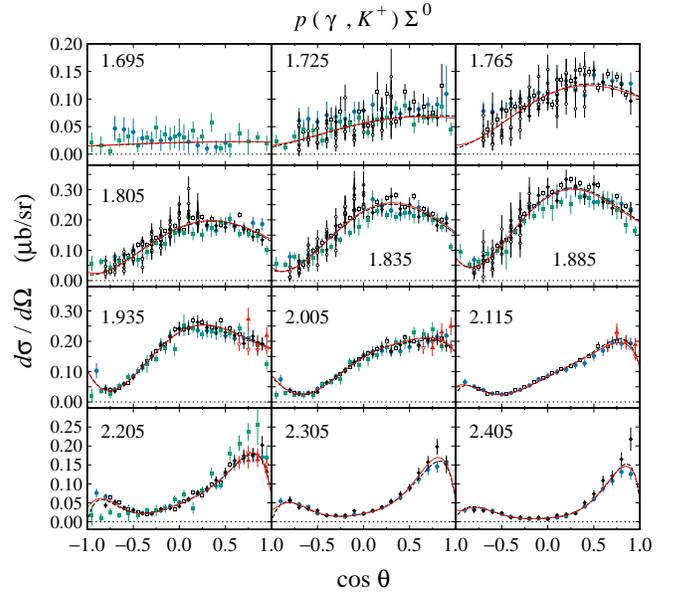}
	\caption{As in Fig.~\ref{fig:dkps_e}, but for angular distribution.
        The corresponding total c.m. energy $W$ is shown in each panel in GeV.}
	\label{fig:dkps_th}
\end{figure}

Comparison between differential cross sections obtained from the two models 
is shown in Figs.~\ref{fig:dkps_e} and \ref{fig:dkps_th}. The energy distribution
of differential cross section shown in Fig.~\ref{fig:dkps_e} reveals that the
difference between the two models is most obvious at $\cos\theta=0.95$. At this
forward angle we can see that the inclusion of the high-spin resonances yields
at least two 
structures in differential cross section at the $W\geq 2.6$ GeV, which can be
traced back to the resonance masses. 

Furthermore, the second peak in differential 
cross section becomes more apparent and much closer to experimental data after 
the inclusion of these resonances. This peak still clearly appears at 
$\cos\theta=0.80$ and 0.65, and quickly disappears as we move to larger 
kaon angles. We have investigated the origin of this second peak and found
that it is due to the $\Delta(2000)F_{35}$ resonance.

In addition, we also observe a third peak at $W\approx 2.3$ GeV, which originates 
from the $N(2290)G_{19}$ state with spin 9/2, has positive parity, and earns a 
status of four-star in the Particle Data Book \cite{pdg}. Interestingly, this state
appears in the $K^+\Sigma^0$ channel after the inclusion of higher nucleon 
resonances. This peak also quickly disappears as we increase the kaon angle,
but appears again in the backward angles. In the latter, both models are in
agreement with each other. 

The angular distributions of differential cross section shown in 
Fig.~\ref{fig:dkps_th} support this finding. In general, the agreement 
with experimental data are similar for the two models. In the case when 
the experimental data from different collaborations are scattered, the 
models try to reproduce their average.

\subsubsection{$K^0\Sigma^+$ Channel}
\begin{figure}[t!]
	\centering
	\includegraphics[scale=0.43]{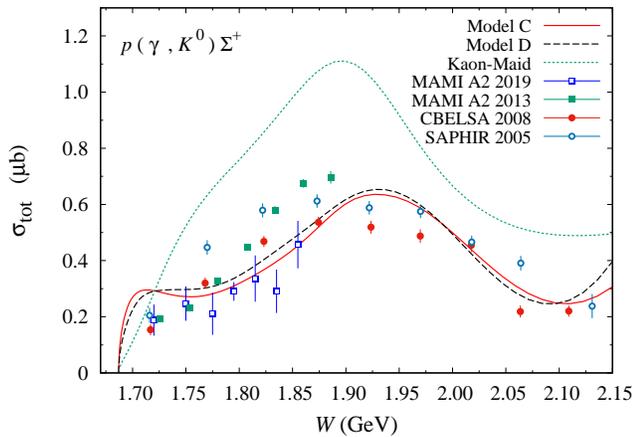}
	\caption{Total cross section for the $\gamma p \to K^0 \Sigma^+$ process
        obtained from the present and previous models. Experimental data are obtained
        from the SAPHIR 2005 \cite{saphir2005}, CBELSA 2008 \cite{cbelsa}, 
        MAMI A2 2013 \cite{mami2013}, and MAMI A2 2019 \cite{mami2018} collaborations.
        All data shown in this figure were not used in the present analysis.}
	\label{fig:total4}
\end{figure}

The $K^0 \Sigma^+$ channel is the last channel measured for the proton target. Although
proton is stable, detection of the neutral kaon and positively charged $\Sigma$
hyperon in the final state is more challenging. Figure~\ref{fig:total4} shows 
the $K^0 \Sigma^+$  total cross sections obtained from both models, as well as
Kaon-Maid for comparison. Obviously, the three models display different shapes
of total cross section, which originate from different nucleon and delta resonances 
used in the models. We note that experimental data for this channel are relatively
scattered, especially in the energy range $1.77\lesssim W\lesssim 1.90$ GeV, where 
the new SAPHIR data are almost 50\% smaller than the older SAPHIR data. Fortunately, 
near the production threshold all data are in agreement with each other and, 
interestingly, closer to the prediction of Kaon-Maid. Nevertheless, at this 
point we still observe that the inclusion of higher-spin resonances improves 
the model prediction. Furthermore, as  shown in Fig.~\ref{fig:total4}, very close 
to the threshold the predictions of the three models are very different. While
the predicted cross section of Kaon-Maid is slowly increasing with energy, 
the calculated cross section of the present model (Model C) is rising steeply 
and reveals the contributions of resonances with masses near 1.7 GeV. We note 
that there are four nucleon resonances and one delta resonance to this end. 
As stated in Ref.~\cite{ks_threshold}, experimental data near the threshold 
region are very crucial to understand the production mechanism and 
related phenomenological applications \cite{Sakinah:2019zbd}. 

Despite the fact that there are limited data available for this reaction, 
the agreement between model calculation and experimental data increases 
significantly with the inclusion of high-spin resonances. As in the case
of previous channels, the inclusion of high-spin resonances leads also
to a number of structures in the total cross sections, as clearly shown
in Fig.~\ref{fig:total4}.

\begin{figure}[t!]
	\centering
	\includegraphics[scale=0.42]{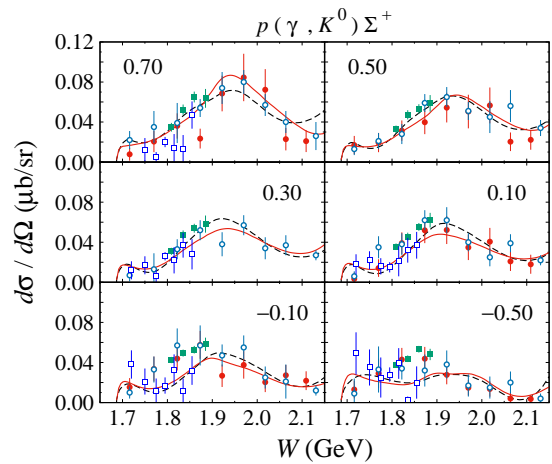}
	\caption{Energy distribution of the $\gamma p \to K^0 \Sigma^+$ differential 
          cross section calculated from Models C and D for different values of $\cos\theta$,
          compared with available experimental data. Notation of the curves and
          experimental data is as in Fig.~\ref{fig:total4}.}
	\label{fig:dk0s_e}
\end{figure}

The limited number of experimental data for the $K^0\Sigma^+$ photoproduction 
clearly impose a strong constraint on the range of validity of our present model. 
Figure~\ref{fig:dk0s_e} obviously shows that the available data are 
relatively scattered, with apparent peak at $W\approx 1.95$ GeV.
This peak is more obvious in the forward regions. Furthermore, in this figure 
we can also observe that the inclusion of high-spin resonances eliminates the small 
structure at $W \approx 1.85$ GeV and emphasizes the contribution of resonances
with $m \approx 1.7$ GeV as in the case of total cross section. More experimental 
data are strongly required, especially at these energy points, to clarify the
effects of the inclusion of high-spin resonances in the present model.

\begin{figure}[t!]
	\centering
	\includegraphics[scale=0.42]{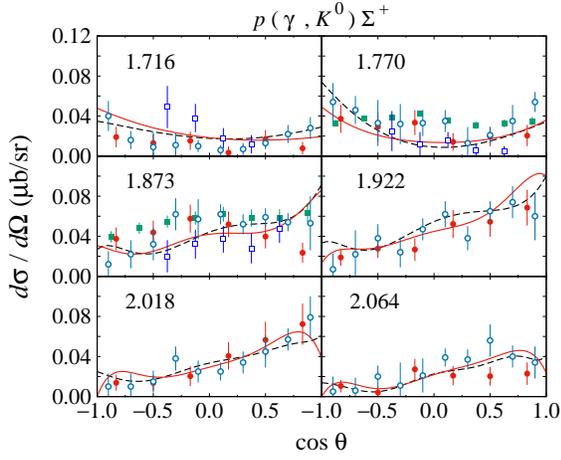}
	\caption{As in Fig.~\ref{fig:dk0s_e}, but for angular distribution.
        The corresponding total c.m. energy $W$ in GeV is shown in each panel.}
	\label{fig:dk0s_th}
\end{figure}

The angular distributions of differential cross section shown in Fig.~\ref{fig:dk0s_th}
clear up the difference between Models C and D, which is visible in the whole
angles covered by experimental data. The improvement of the model after including 
the high-spin resonances is relatively unclear due to the scattered experimental 
data. In general, the inclusion of the high-spin resonances slightly improves the 
agreement between model calculation and experimental data. Both Figs.~\ref{fig:dk0s_e}
and \ref{fig:dk0s_th} indicate that the present model prefers the new MAMI A2 2019 
data set \cite{mami2018} in the energy range $1.8 \lesssim W \lesssim 1.9$ GeV, where
experimental data from different collaborations are significantly scattered.

\subsubsection{$K^+\Sigma^-$ Channel}
\begin{figure}[b!]
	\centering
	\includegraphics[scale=0.40]{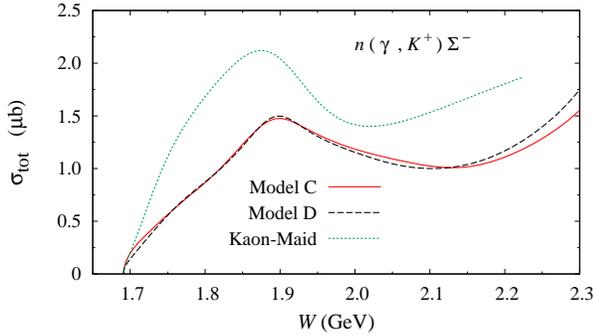}
	\caption{Calculated otal cross section of $\gamma n \to K^+ \Sigma^-$ process.}
	\label{fig:total5}
\end{figure}

The $K^+ \Sigma^-$ photoproduction channel has nearly 300 experimental data points
in the form of differential cross section \cite{sigleps2006,clas2010sig} and photon 
asymmetry \cite{sigleps2006}. These data were included in the fitting process. 
Thus, the total cross sections shown in Fig.~\ref{fig:total5} are pure prediction
and cannot be compared with experimental measurement. Nevertheless, we still can 
see the small effect of the higher-spin resonances near the $K^+ \Sigma^-$ threshold, 
as in the case of $K^0\Sigma^+$ channel, and at high energy $W\gtrsim 2.1$ GeV where
no experimental data are available to constrain the model. Otherwise, both models C 
and D show similar trend.

\begin{figure}[h!]
	\centering
	\includegraphics[scale=0.5]{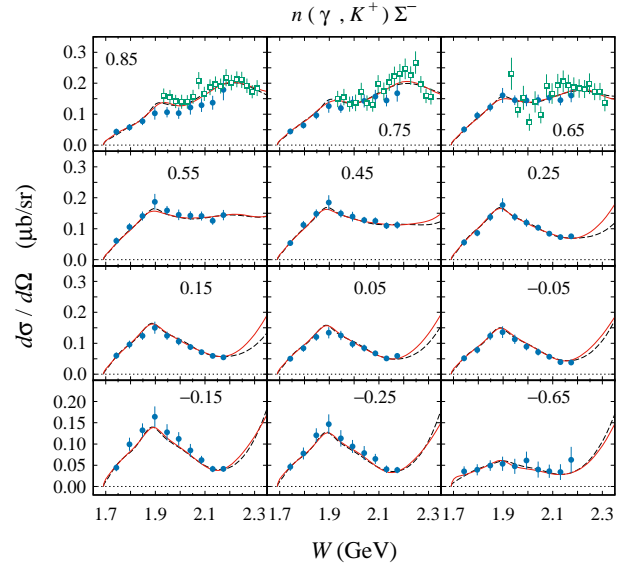}
	\caption{Energy distribution of the $\gamma n \to K^+ \Sigma^-$ 
          differential cross section. Notation of the curves is as in 
          Fig.~\ref{fig:total5}. Experimental data are obtained from the
          LEPS 2006 \cite{sigleps2006} and CLAS 2010 \cite{clas2010sig}
          collaborations.}
	\label{fig:dkpsm_e}
\end{figure}

The calculated  $K^+ \Sigma^-$ differential cross sections obtained from both models are compared
with experimental data in Fig.~\ref{fig:dkpsm_e}. This figure reveals that the origin
of the structures shown in the total cross section. Near the threshold the peak appears
in the whole angular distribution, whereas the difference between the two models 
at high energy originates from the backward angles. It is interesting to note that
at high energy the inclusion of the high-spin resonances slightly increases the cross 
section, but in the backward region this situation dramatically changes (see the panel
with $\cos\theta=-0.65$).

\begin{figure}[h!]
	\centering
	\includegraphics[scale=0.53]{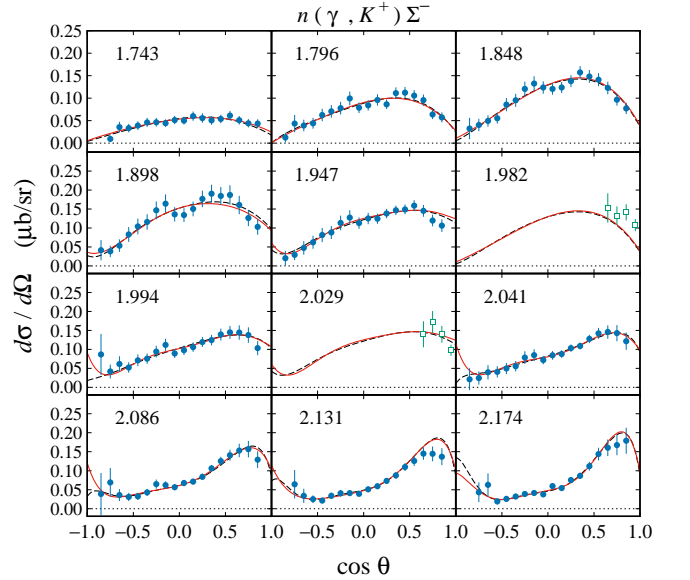}
	\caption{As in Fig.~\ref{fig:dkpsm_e}, but for angular distribution.}
	\label{fig:dkpsm_th}
\end{figure}

For completeness, we have also checked the angular distribution of $K^+ \Sigma^-$ differential
cross section. Figure~\ref{fig:dkpsm_th} shows the comparison between both models and experimental 
data. Similar to Fig.~\ref{fig:dkpsm_e}, we can see that the difference between models C and D 
is also small, except at high energy, i.e, at $W = 2.174$ GeV and backward angle. 

\subsubsection{$K^0\Sigma^0$ channel}
\begin{figure}[t!]
	\centering
	\includegraphics[scale=0.43]{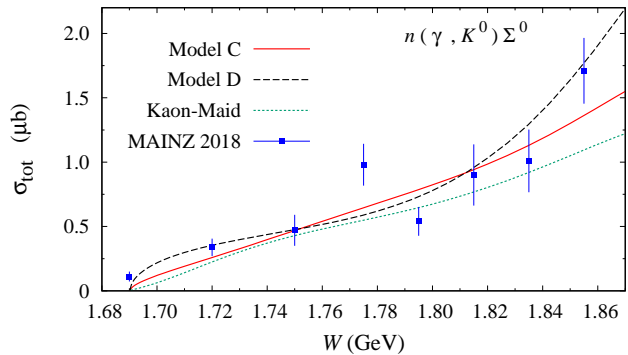}
	\caption{Total cross section of $\gamma n \to K^0 \Sigma^0$ process.
         Experimental data \cite{mami2018} shown in this figure 
         were not included in the fitting process of the present model.}
	\label{fig:total6}
\end{figure}

The $K^0 \Sigma^0$ channel has very limited experimental data. They were obtained 
by the MAMI A2 2018 collaboration by measuring photoproduction of neutral kaon on 
a deuteron target. Although the number is very limited, the existence of experimental
data in this channel significantly helps to constrain the prediction of the present
model. Figure~\ref{fig:total6} 
shows the comparison between calculated total cross sections obtained from previous
and present models and experimental data. It can be seen that the inclusion of
the high-spin resonances improves the model, although the cross section trend is
relatively well reproduced by the three models. The importance of the nucleon and
delta resonances with $m\approx 1.7$ GeV is slightly shown by Model D near the
production threshold. 

As shown in Fig.~\ref{fig:total6} the total cross sections increase monotonically
with increasing energy. In fact, the predicted total cross section of the present 
model is more than 3 $\mu$b at $W \approx 2$ GeV, which seems to be unrealistic if 
we compare it with those of the neutral kaon productions shown in 
Figs.~\ref{fig:k0ltotal} and \ref{fig:total4}. Thus, total cross section data up to
2.5 GeV are very important to this end.

\begin{figure}[t!]
	\centering
	\includegraphics[scale=0.4]{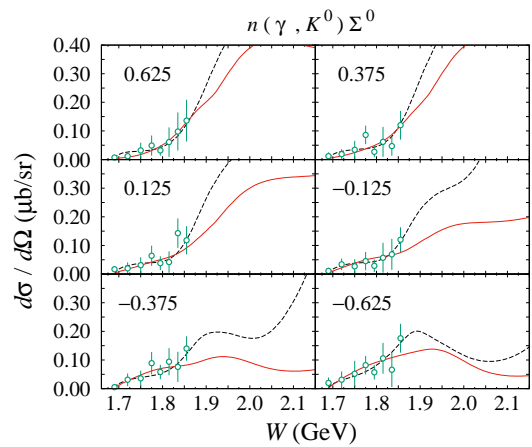}
	\caption{Energy distribution of the $\gamma n \to K^0 \Sigma^0$ differential 
          cross section. Notation for the curves and experimental data is as in 
          Fig.~\ref{fig:total6}. }
	\label{fig:dk0s0_e}
\end{figure}

\begin{figure}[t!]
	\centering
	\includegraphics[scale=0.4]{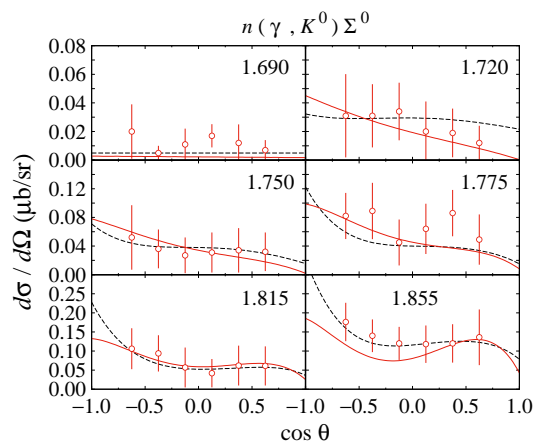}
	\caption{As in Fig.~\ref{fig:dk0s0_e}, but for angular distribution.}
	\label{fig:dk0s0_th}
\end{figure}

Figures~\ref{fig:dk0s0_e} and \ref{fig:dk0s0_th} compare the calculated differential 
cross sections obtained from the previous and present models with experimental data. Figure~\ref{fig:dk0s0_e} shows that
models C and D start to differ at $W\approx 1.86$ GeV, where no experimental data are
available to constrain them. The inclusion of the high-spin resonances in this channel 
improves the cross section divergence, which is urgently required in forward regions.
Interestingly, the two models predict a resonance structure above this energy point, 
albeit with different positions. Certainly, experimental data in the energy range 
$1.8\lesssim W \lesssim 2.2$ GeV are very important to determine which resonance is 
responsible for this structure.

The angular distributions of differential cross section shown in Fig.~\ref{fig:dk0s0_th} 
reveals that the cross section of the $K^0\Sigma^0$ photoproduction has backward-peaking
behavior. This indicates the dominance of $u$-channel in this process, which is easily
understood from the fact that this process does not have a $t$-channel in the Born terms
since a neutral kaon cannot interact with real photon.

\subsection{Extracted Resonance Properties}

Having investigated the effect of high-spin resonances on our models, we are ready 
to discuss the resonance properties, i.e., their masses and widths, at their pole 
positions, before and after including the high-spin resonances. Note that during 
the fitting process we allowed the resonance masses and widths to vary within the
estimated error bars of PDG.
In Table~\ref{tab:pole} 
we show the resonance masses and widths evaluated at their pole positions extracted 
from the present and previous works, compared with those listed by PDG~\cite{pdg}. 
Note that for the $N(2600)I_{1,11}$, $N(2700)K_{1,13}$, $\Delta(2750)I_{3,13}$ and 
$\Delta(2950)K_{3,15}$ resonances, the PDG does not have any information yet. 
Therefore, in this case, the result shown in Table~\ref{tab:pole} provides the first estimate 
for these resonances.

\begin{table*}[!]
    \centering
    \caption {Masses and widths of nucleon and $\Delta$ resonances evaluated at the pole position 
      in MeV, obtained from the present work (Models A and C), previous works (Models B~\cite{Mart:2019mtq} 
      and D~\cite{samson_ksigma}) and PDG~\cite{pdg}. The status of resonances is due to the 
      PDG~\cite{pdg}. \label{tab:pole}}
    \begin{ruledtabular}
    \begin{tabular}{lccccccccccc}
      Resonances & Status & \multicolumn{2}{c}{PDG} & \multicolumn{2}{c}{Model A} & \multicolumn{2}{c}{Model B } & 
      \multicolumn{2}{c}{Model C} & \multicolumn{2}{c}{Model D} \\	
      & & $m_{\rm{pole}}$ & $\Gamma_{\rm{pole}}$ & $m_{\rm{pole}}$ & $\Gamma_{\rm{pole}}$ & $m_{\rm{pole}}$ & 
      $\Gamma_{\rm{pole}}$  & $m_{\rm{pole}}$ & $\Gamma_{\rm{pole}}$  & $m_{\rm{pole}}$ & $\Gamma_{\rm{pole}}$\\
      \hline\\[-2ex]
      $N(1440)P_{11}$         &**** & 1370 $\pm$ 10 & 175 $\pm$ 15 & 1305 $\pm$ 27.9 & 173 $\pm$ 46.0 & 1305 & 173               & 1355 $\pm$ 5.2  & 208 $\pm$ 7.2  & 1324 & 188 \\
      $N(1520)D_{13}$         &**** & 1510 $\pm$ 5  & 110 $^{+10}_{-5}$ & 1488 $\pm$ 3.5  & 112 $\pm$ 10.8 & 1495 & 101          & 1487 $\pm$ 4.4  & 112 $\pm$ 11.7 & 1489 & 100 \\
      $N(1535)S_{11}$         &**** & 1510 $\pm$ 10 & 130 $\pm$ 20               & 1508 $\pm$ 15.0 & 169 $\pm$ 26.0 & 1475 & 162               & 1474 $\pm$ 15.2 & 225 $\pm$ 17.3 & 1530 & 129 \\
      $N(1650)S_{11}$         &**** & 1655 $\pm$ 15 & 135 $\pm$ 35               & 1599 $\pm$ 1.5  & 230 $\pm$ 5.0  & 1612 & 232               & 1600 $\pm$ 18.1 & 230 $\pm$ 4.7  & 1664 & 176 \\
      $N(1675)D_{15}$         &**** & 1660 $\pm$ 5  & 135 $^{+15}_{-10}$         & 1639 $\pm$ 0.0  & 147 $\pm$ 0.2  & 1640 & 147         & 1629 $\pm$ 2.7  & 164 $\pm$ 14.1 & 1643 & 136 \\
      $N(1680)F_{15}$         &**** & 1675 $^{+5}_{-10}$ & 120 $^{+15}_{-10}$    & 1644 $\pm$ 0.9  & 119 $\pm$ 0.6  & 1651 & 123    & 1653 $\pm$ 2.9  & 121 $\pm$ 7.6  & 1667 & 98\\
      $N(1700)D_{13}$         &***  & 1700 $\pm$ 50 & 200 $\pm$ 100              & 1668 $\pm$ 0.3  & 155 $\pm$ 0.2  & 1692 & 158              & 1637 $\pm$ 0.6  & 183 $\pm$ 0.6  & 1630 & 111 \\
      $N(1710)P_{11}$         &**** & 1700 $\pm$ 20 & 120 $\pm$ 40               & 1658 $\pm$ 0.9  & 120 $\pm$ 2.9  & 1657 & 175               & 1676 $\pm$ 0.8  & 71  $\pm$ 3.0  & 1705 & 47\\
      $N(1720)P_{13}$         &**** & 1675 $\pm$ 15 & 250 $^{+150}_{-100}$       & 1632 $\pm$ 0.1  & 220 $\pm$ 0.4  & 1648 & 193       & 1578 $\pm$ 0.7  & 242 $\pm$ 1.5  & 1665 & 300 \\
      $N(1860)F_{15}$         &**   & 1830 $^{+120}_{-60}$ & 250 $^{+150}_{-50}$ & 1874 $\pm$ 0.2  & 226 $\pm$ 0.4  & 1862 & 217 & 1836 $\pm$ 0.4  & 241 $\pm$ 0.7  & 1787 & 156 \\
      $N(1875)D_{13}$         &***  & 1900 $\pm$ 50 & 160 $\pm$ 60               & 1840 $\pm$ 0.2  & 230 $\pm$ 1.0  & 1815 & 222               & 1765 $\pm$ 4.9  & 223 $\pm$ 0.7  & 1757 & 219 \\
      $N(1880)P_{11}$         &***  & 1860 $\pm$ 40 & 230 $\pm$ 50               & 1753 $\pm$ 1.7  & 325 $\pm$ 0.7  & 1785 & 298               & 1836 $\pm$ 4.8  & 278 $\pm$ 3.2  & 1831 & 166 \\
      $N(1895)S_{11}$         &**** & 1910 $\pm$ 20 & 110 $\pm$ 30               & 1874 $\pm$ 0.2  & 299 $\pm$ 0.4  & 1876 & 224               & 1848 $\pm$ 0.1  & 320 $\pm$ 0.4  & 1893 & 90 \\
      $N(1900)P_{13}$         &**** & 1920 $\pm$ 20 & 150 $\pm$ 50               & 1846 $\pm$ 0.2  & 265 $\pm$ 0.4  & 1865 & 256               & 1874 $\pm$ 0.7  & 186 $\pm$ 0.8  & 1899 & 239 \\
      $N(1990)F_{17}$         &**   & 2030 $\pm$ 65 & 240 $\pm$ 60               & 1935 $\pm$ 0.8  & 249 $\pm$ 1.3  & 1867 & 240               & 1916 $\pm$ 0.5  & 225 $\pm$ 1.2  & 2044 & 273 \\
      $N(2000)F_{15}$         &**   & 2030 $\pm$ 40 & 380 $\pm$ 60               & 1858 $\pm$ 0.2  & 264 $\pm$ 0.5  & 1932 & 273               & 1898 $\pm$ 1.2  & 260 $\pm$ 2.6  & 1978 & 232 \\
      $N(2060)D_{15}$         &***  & 2070 $^{+60}_{-50}$ & 400 $^{+30}_{-50}$   & 1950 $\pm$ 0.7  & 400 $\pm$ 0.3  & 1856 & 341    & 1951 $\pm$ 1.2  & 401 $\pm$ 1.1  & 1968 & 334 \\
      $N(2120)D_{13}$         &***  & 2100 $\pm$ 50 & 280 $\pm$ 60               & 1963 $\pm$ 0.7  & 354 $\pm$ 0.1  & 1884 & 354                & 1957 $\pm$ 0.7  & 385 $\pm$ 1.5  & 2029 & 274 \\
      $N(2190)G_{17}$         &**** & 2100 $\pm$ 50 & 400 $\pm$ 100              & 2014 $\pm$ 0.6  & 241 $\pm$ 0.3  & 2025 & 244               & 2059 $\pm$ 30.7 & 256 $\pm$ 64.4 & 2142 & 211 \\
      $N(2220)H_{19}$         &**** & 2170  $^{+30}_{-40}$ & 400 $^{+80}_{-40}$  & 2047 $\pm$ 0.4  & 220 $\pm$ 2.0  & 2020 & 228  & 2114 $\pm$ 18.4 & 240 $\pm$ 53.0 & 2131 & 202 \\
      $N(2250)G_{19}$         &**** & 2200 $\pm$ 50 & 420 $^{+80}_{-80}$         & 2138 $\pm$ 0.5  & 284 $\pm$ 4.9  & 2085 & 265         & 2138 $\pm$ 2.2  & 285 $\pm$ 4.6  & 2193 & 219 \\
      $N(2600)I_{1,11}$       &***  & $\cdots$ & $\cdots$ & 2318 $\pm$ 1.9  & 272 $\pm$ 33.9 & $\cdots$ & $\cdots$ & 2389 $\pm$ 11.4 & 310 $\pm$ 34.9 & $\cdots$ & $\cdots$ \\
      $N(2700)K_{1,13}$       &**   & $\cdots$ & $\cdots$ & 2393 $\pm$ 3.4  & 247 $\pm$ 18.0 & $\cdots$ & $\cdots$ & 2457 $\pm$ 17.7 & 253 $\pm$ 41.8 & $\cdots$ & $\cdots$ \\
      $\Delta(1232)P_{33}$    &**** & 1210 $\pm$ 1 & 100 $\pm$ 2 & $\cdots$ & $\cdots$ & $\cdots$ & $\cdots$     & 1209 $\pm$ 1.7  & 82  $\pm$ 5.2  & 1205 & 82 \\
      $\Delta(1600)P_{33}$    &**** & 1510 $\pm$ 50 & 270 $\pm$ 70 & $\cdots$ & $\cdots$ & $\cdots$ & $\cdots$ & 1444 $\pm$ 64.3 & 170 $\pm$ 52.5 & 1457 & 168 \\
      $\Delta(1620)S_{31}$    &**** & 1600 $\pm$ 10 & 120 $\pm$ 20 & $\cdots$ & $\cdots$ & $\cdots$ & $\cdots$ & 1565 $\pm$ 15.6 & 140 $\pm$ 20.2 & 1598 & 152 \\
      $\Delta(1700)D_{33}$    &**** & 1665 $\pm$ 25 & 250 $\pm$ 50 & $\cdots$ & $\cdots$ & $\cdots$ & $\cdots$ & 1616 $\pm$ 3.6  & 221 $\pm$ 1.2  & 1646 & 161 \\
      $\Delta(1900)S_{31}$    &***  & 1865 $\pm$ 35 & 240 $\pm$ 60 & $\cdots$ & $\cdots$ & $\cdots$ & $\cdots$ & 1760 $\pm$ 0.3  & 375 $\pm$ 1.0  & 1938 & 330 \\
      $\Delta(1905)F_{35}$    &**** & 1800 $\pm$ 30 & 300 $\pm$ 40 & $\cdots$ & $\cdots$ & $\cdots$ & $\cdots$ & 1772 $\pm$ 4.7  & 228 $\pm$ 15.3 & 1797 & 212 \\
      $\Delta(1910)P_{31}$    &**** & 1860 $\pm$ 30 & 300 $\pm$ 100 & $\cdots$ & $\cdots$ & $\cdots$ & $\cdots$ & 1836 $\pm$ 0.4  & 351 $\pm$ 0.9  & 1859 & 317 \\
      $\Delta(1920)P_{33}$    &***  & 1900 $\pm$ 50 & 300 $\pm$ 100 & $\cdots$ & $\cdots$ & $\cdots$ & $\cdots$ & 1758 $\pm$ 4.0  & 281 $\pm$ 7.7  & 1893 & 193 \\
      $\Delta(1930)D_{35}$    &***  & 1880 $\pm$ 40 & 280 $\pm$ 50 & $\cdots$ & $\cdots$ & $\cdots$ & $\cdots$ & 1850 $\pm$ 0.2  & 314 $\pm$ 0.7  & 1933 & 199 \\
      $\Delta(1940)D_{33}$    &**   & 1950 $\pm$ 100 & 350 $\pm$ 150 & $\cdots$ & $\cdots$ & $\cdots$ & $\cdots$ & 1878 $\pm$ 0.2  & 334 $\pm$ 1.1  & 1880 & 349 \\
      $\Delta(1950)F_{37}$    &**** & 1880 $\pm$ 10 & 240 $\pm$ 20 & $\cdots$ & $\cdots$ & $\cdots$ & $\cdots$ & 1835 $\pm$ 2.3  & 208 $\pm$ 5.7  & 1848 & 208 \\
      $\Delta(2000)F_{35}$    &**   & 2150 $\pm$ 100 & 350 $\pm$ 100 & $\cdots$ & $\cdots$ & $\cdots$ & $\cdots$ & 1844 $\pm$ 16.8 & 275 $\pm$ 42.4 & 2081 & 328 \\
      $\Delta(2300)H_{39}$    &**   & 2370 $\pm$ 80 & 420 $\pm$ 160 & $\cdots$ & $\cdots$ & $\cdots$ & $\cdots$ & 2318 $\pm$ 4.7  & 288 $\pm$ 4.3  & 2341 & 200 \\
      $\Delta(2400)G_{39}$    &**   & 2260 $\pm$ 60 & 320 $\pm$ 160 & $\cdots$ & $\cdots$ & $\cdots$ & $\cdots$ & 2282 $\pm$ 1.2  & 386 $\pm$ 4.1  & 2409 & 329 \\
      $\Delta(2420)H_{3,11}$  &**** & 2400 $\pm$ 100 & 450 $\pm$ 100 & $\cdots$ & $\cdots$ & $\cdots$ & $\cdots$ & 2302 $\pm$ 60.4 & 288 $\pm$ 118.3& $\cdots$ & $\cdots$ \\
      $\Delta(2750)I_{3,13}$  &**   & $\cdots$ & $\cdots$ & $\cdots$ & $\cdots$ & $\cdots$ & $\cdots$ & 2454 $\pm$ 28.3 & 309 $\pm$ 64.3& $\cdots$ & $\cdots$ \\   	    	   	 
      $\Delta(2950)K_{3,15}$  &**   & $\cdots$ & $\cdots$ & $\cdots$ & $\cdots$ & $\cdots$ & $\cdots$ & 2572 $\pm$ 33.6 & 297 $\pm$ 88.2& $\cdots$ & $\cdots$ \\   	    	   	 
    \end{tabular}
    \end{ruledtabular}
\end{table*}

\begin{table*}[!h] 
	\centering
	\caption {Breit-Wigner mass and width of nucleon and $\Delta$ resonances in MeV from present work (Models A and C) previous works (Models B \cite{Mart:2019mtq} and D \cite{samson_ksigma}) and PDG \cite{pdg}.\label{tab:breit-wigner} }
	\hspace*{-2.0cm}
        \begin{ruledtabular}
	\begin{tabular}{lcccccccccc}
          Resonances &\multicolumn{2}{c}{PDG} & \multicolumn{2}{c}{Model A ($K\Lambda$)} & 
          \multicolumn{2}{c}{Model B ($K\Lambda$)} & \multicolumn{2}{c}{Model C ($K\Sigma$)} & \multicolumn{2}{c}{Model D ($K\Sigma$)} \\	
		& Mass & Width & Mass & Width & Mass & Width  & Mass & Width  & Mass & Width\\
		\hline\\[-2ex]
		$N(1440)P_{11}$ & 1440$\pm$30	     & 350$\pm$100	      & 1410 $\pm$ 49  & 450 $\pm $41.3 & 1410 & 450  & 1470 $\pm$ 8.0  & 450 $\pm$ 7.6   & 1420 & 450 \\
		$N(1520)D_{13}$ & 1515$\pm$5	     & 110$\pm$10	      & 1520 $\pm$ 2.1 & 120 $\pm $12.3 & 1520 & 100  & 1519 $\pm$ 0.9  & 120 $\pm$ 13.3  & 1510 & 125 \\
		$N(1535)S_{11}$ & 1530$\pm$15	     & 150$\pm$25	      & 1545 $\pm$ 3.7 & 125 $\pm $48.5 & 1545 & 175  & 1545 $\pm$ 27.7 & 175 $\pm$ 32.7  & 1545 & 125 \\
		$N(1650)S_{11}$ & 1650$\pm$15	     & 125$\pm$25	      & 1545 $\pm$ 2.5 & 150 $\pm $0.8  & 1645 & 159  & 1635 $\pm$ 29.4 & 150 $\pm$ 9.8   & 1670 & 170 \\
		$N(1675)D_{15}$ & 1675$\pm$15	     & 145$\pm$15	      & 1635 $\pm$ 0.1 & 130 $\pm $0.3  & 1680 & 130  & 1680 $\pm$ 9.6  & 160 $\pm$ 21.2  & 1670 & 165 \\
		$N(1680)F_{15}$ & 1685$\pm$5	     & $120^{+10}_{-5}$      & 1680 $\pm$ 1.5 & 115 $\pm $0.7  & 1690 & 120  & 1690 $\pm$ 1.0  & 115 $\pm$ 9.9   & 1686 & 120 \\
		$N(1700)D_{13}$ & $1720^{+80}_{-70}$ & 200$\pm$100	      & 1712 $\pm$ 0.5 & 134 $\pm $0.5  & 1731 & 102  & 1706 $\pm$ 1.0  & 201 $\pm$ 0.9   & 1650 & 129 \\
		$N(1710)P_{11}$ & 1710$\pm$30	     & 140$\pm$60	      & 1708 $\pm$ 0.5 & 182 $\pm $3.0  & 1733 & 1250 & 1697 $\pm$ 0.4  & 80  $\pm$ 2.8   & 1710 & 50 \\
		$N(1720)P_{13}$ & $1720^{+30}_{-40}$ & $250^{+150}_{-100}$   & 1703 $\pm$ 0.5 & 208 $\pm $0.6  & 1700 & 2189 & 1687 $\pm$ 0.0  & 340 $\pm$ 2.3   & 1750 & 400 \\
		$N(1860)F_{15}$ & 1928$\pm$21	     & 376$\pm$58	      & 1980 $\pm$ 0.7 & 235 $\pm $0.8  & 1960 & 220  & 1971 $\pm$ 1.0  & 337 $\pm$ 1.2   & 1829 & 220 \\
		$N(1875)D_{13}$ & $1875^{+45}_{-25}$ & $200^{+150}_{-80}$    & 1918 $\pm$ 1.4 & 177 $\pm $2.2  & 1858 & 180  & 1850 $\pm$ 8.7  & 213 $\pm$ 0.9   & 1820 & 320 \\
		$N(1880)P_{11}$ & 1880$\pm$50	     & 300$\pm$100	      & 1930 $\pm$ 4.4 & 400 $\pm $1.2  & 1915 & 280  & 1930 $\pm$ 9.6  & 200 $\pm$ 7.7   & 1856 & 180 \\
		$N(1895)S_{11}$ & 1895$\pm$25	     & $120^{+80}_{-40}$     & 1903 $\pm$ 0.7 & 142 $\pm $1.2  & 1893 & 106  & 1884 $\pm$ 0.7  & 157 $\pm$ 1.4   & 1893 & 90 \\
		$N(1900)P_{13}$ & 1920$\pm$30	     & $200^{+120}_{-100}$   & 1920 $\pm$ 0.6 & 169 $\pm $1.1  & 1930 & 151  & 1907 $\pm$ 1.6  & 100 $\pm$ 1.9   & 1930 & 250 \\
		$N(1990)F_{17}$ & $2020^{+80}_{-70}$ & 300$\pm$100	      & 2057 $\pm$ 2.9 & 245 $\pm $2.7  & 1995 & 265  & 2013 $\pm$ 1.5  & 200 $\pm$ 2.5   & 2125 & 400 \\
		$N(2000)F_{15}$ & 2060$\pm$30	     & 390$\pm$55	      & 2030 $\pm$ 0.4 & 445 $\pm $0.9  & 2090 & 338  & 2046 $\pm$ 0.9  & 335 $\pm$ 4.8   & 2044 & 335 \\
		$N(2060)D_{15}$ & $2100^{+100}_{-70}$& $400^{+50}_{-100}$    & 2200 $\pm$ 2.2 & 450 $\pm $0.8  & 2060 & 450  & 2200 $\pm$ 2.7  & 450 $\pm$ 2.7   & 2060 & 450 \\
		$N(2120)D_{13}$ & $2120^{+40}_{-60}$ & $300^{+60}_{-40}$     & 2126 $\pm$ 1.4 & 275 $\pm $0.3  & 2075 & 375  & 2160 $\pm$ 1.6  & 345 $\pm$ 4.2   & 2075 & 305 \\
		$N(2190)G_{17}$ & 2180$\pm$40	     & 400$\pm$100	      & 2159 $\pm$ 1.8 & 300 $\pm $0.5  & 2181 &300   & 2216 $\pm$ 5.1  & 300 $\pm$ 127.8 & 2200 & 300 \\
		$N(2220)H_{19}$ & 2250$\pm$50	     & $400^{+100}_{-50}$    & 2200 $\pm$ 2.5 & 350 $\pm $3.1  & 2200 & 500  & 2284 $\pm$ 24.2 & 350 $\pm$ 92.7  & 2204 & 369 \\
		$N(2250)G_{19}$ & $2280^{+40}_{-30}$ & $500^{+100}_{-200}$   & 2320 $\pm$ 1.8 & 300 $\pm $11.2 & 2283 & 300  & 2320 $\pm$ 2.2  & 300 $\pm$ 10.4  & 2250 & 300 \\
                $N(2600)I_{1,11}$ & $2600^{+150}_{-50}$& 650$\pm$150	      & 2573 $\pm$ 6.9 & 500 $\pm $61.8 & $\cdots$ & $\cdots$       & 2750 $\pm$ 7.8  & 800 $\pm$ 59.1  & $\cdots$ & $\cdots$ \\
	        $N(2700)K_{1,13}$ & 2612$\pm$45        & 350$\pm$50            & 2621 $\pm$ 3.3 & 400 $\pm $17.8 & $\cdots$ & $\cdots$       & 2675 $\pm$ 9.6  & 300 $\pm$ 89.9  & $\cdots$ & $\cdots$ \\
		$\Delta(1232)P_{33}$   & 1232$\pm$2	    & 117$\pm$3	   & $\cdots$ & $\cdots$ & $\cdots$ & $\cdots$ & 1234 $\pm$ 2.8  & 114 $\pm$ 4.0    & 1230 & 120 \\
		$\Delta(1600)P_{33}$   & 1570$\pm$70	    & 250$\pm$50	   & $\cdots$ & $\cdots$ & $\cdots$ & $\cdots$ & 1500 $\pm$ 99   & 200 $\pm$ 68.6   & 1500 & 220 \\
		$\Delta(1620)S_{31}$   & 1610$\pm$20	    & 130$\pm$20	   & $\cdots$ & $\cdots$ & $\cdots$ & $\cdots$ & 1590 $\pm$ 28.8 & 110 $\pm$ 30.3	 & 1600 & 150 \\
		$\Delta(1700)D_{33}$   & 1710$\pm$20	    & 300$\pm$800	   & $\cdots$ & $\cdots$ & $\cdots$ & $\cdots$ & 1730 $\pm$ 5.9  & 355 $\pm$ 1.2	 & 1686 & 213 \\
		$\Delta(1900)S_{31}$   & $1860^{+80}_{-20}$ & 250$\pm$70	   & $\cdots$ & $\cdots$ & $\cdots$ & $\cdots$ & 1920 $\pm$ 1.2  & 320 $\pm$ 2.7	 & 1920 & 325 \\
		$\Delta(1905)F_{35}$   & $1880^{+30}_{-25}$ & $330^{+70}_{-60}$    & $\cdots$ & $\cdots$ & $\cdots$ & $\cdots$ & 1910 $\pm$ 6.7  & 400 $\pm$ 21.9	 & 1878 & 400 \\
		$\Delta(1910)P_{31}$   & 1900$\pm$50	    & 300$\pm$100	   & $\cdots$ & $\cdots$ & $\cdots$ & $\cdots$ & 1950 $\pm$ 1.0  & 242 $\pm$ 2.5	 & 1910 & 340 \\
		$\Delta(1920)P_{33}$   & 1920$\pm$50	    & 300$\pm$60	   & $\cdots$ & $\cdots$ & $\cdots$ & $\cdots$ & 1970 $\pm$ 1.2  & 360 $\pm$ 14.5	 & 1908 & 195 \\
		$\Delta(1930)D_{35}$   & 1950$\pm$50	    & 300$\pm$100	   & $\cdots$ & $\cdots$ & $\cdots$ & $\cdots$ & 2000 $\pm$ 0.6  & 313 $\pm$ 1.5	 & 1963 & 220 \\
		$\Delta(1940)D_{33}$   & 2000$\pm$60	    & 400$\pm$100	   & $\cdots$ & $\cdots$ & $\cdots$ & $\cdots$ & 2044 $\pm$ 1.3  & 330 $\pm$ 2.7	 & 1994 & 520 \\
		$\Delta(1950)F_{37}$   & 1930$\pm$15	    & 285$\pm$50.0	   & $\cdots$ & $\cdots$ & $\cdots$ & $\cdots$ & 1934 $\pm$ 2.2  & 235 $\pm$ 9.7	 & 1915 & 335 \\
		$\Delta(2000)F_{35}$   & 2015$\pm$24	    & 500$\pm$52	   & $\cdots$ & $\cdots$ & $\cdots$ & $\cdots$ & 2039 $\pm$ 6.8  & 552 $\pm$ 65.6	 & 2192 & 525 \\
		$\Delta(2300)H_{39}$   & 2400$\pm$125	    & 425$\pm$150	   & $\cdots$ & $\cdots$ & $\cdots$ & $\cdots$ & 2525 $\pm$ 12   & 275 $\pm$ 10.6	 & 2393 & 275 \\
		$\Delta(2400)G_{39}$   & 2643$\pm$141	    & 895$\pm$432	   & $\cdots$ & $\cdots$ & $\cdots$ & $\cdots$ & 2641 $\pm$ 4.2  & 544 $\pm$ 10.1	 & 2504 & 463 \\
		$\Delta(2420)H_{3,11}$ & 2300$\pm$150	    & 500$\pm$200	   & $\cdots$ & $\cdots$ & $\cdots$ & $\cdots$ & 2515 $\pm$ 12.5 & 300 $\pm$ 281.3 & $\cdots$ & $\cdots$ \\
		$\Delta(2750)I_{3,13}$ & 2794$\pm$80	    & 350$\pm$100	   & $\cdots$ & $\cdots$ & $\cdots$ & $\cdots$ & 2760 $\pm$ 2.3  & 450 $\pm$ 141.0 & $\cdots$ & $\cdots$ \\
		$\Delta(2950)K_{3,15}$ & 2990$\pm$100       & 330$\pm$100          & $\cdots$ & $\cdots$ & $\cdots$ & $\cdots$ & 2890 $\pm$ 129.2& 430 $\pm$ 193.3 & $\cdots$ & $\cdots$ \\
	\end{tabular}
  \end{ruledtabular}
\end{table*}

In the $K\Lambda$ photoproduction the agreement between the extracted properties
of nucleon resonances and those of PDG is in general 
from fair to good. The same situation is
also seen in the $K\Sigma$ one. Good agreement with the PDG values is observed in 
the case of the $N(1520)D_{13}$ and $\Delta(1232)P_{33}$ resonances. Certain resonances with 
low rating status, e.g., the $N(1875)D_{13}$, $N(2000)F_{15}$ and $\Delta(1900)S_{31}$ states, 
show notable deviation from the PDG value. This indicates that the less-established 
resonances tend to produce the resonance properties that significantly deviates from 
the PDG values. From Table~\ref{tab:pole} it is also apparent that the nucleon properties
obtained from Model A have a better agreement with the PDG values, in contrast to Model B, 
especially for the resonances whose masses are less than 2000 MeV, as well as for 
the $N(2220)H_{19}$ and $N(2250)G_{19}$ states.  Furthermore, we can also observe that
Model D seems to produce higher extracted masses than Model C. On the other hand, 
the calculated widths obtained from Model C are found to be closer to the PDG values, 
except for the $N(2190)G_{17}$, $N(2220)H_{19}$ and $N(2250)G_{19}$ resonances.

\begin{figure}[!]
	\centering
	\includegraphics[scale=0.39]{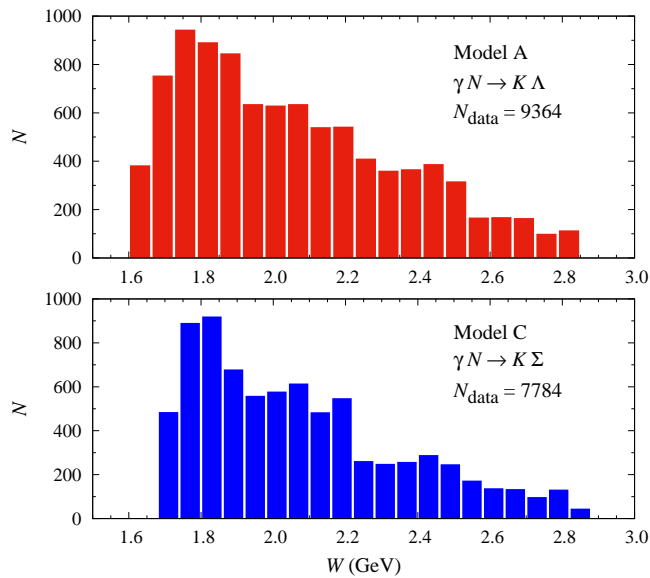}
	\caption{Energy distributions of the experimental data used in the fitting
                 process of Model A ($K\Lambda$ photoproduction) and model C
                 ($K\Sigma$ photoproduction).}
	\label{fig:data_dist}
\end{figure}

The difference between the mass of $N(2700)K_{1,13}$ resonance 
extracted from the $K\Lambda$ and $K\Sigma$ channels seems to be large, i.e., nearly
100 MeV. Furthermore, the extracted masses are smaller than the Breit-Wigner
ones, as can be seen in Table~\ref{tab:pole}. Nevertheless, the masses and widths of
the two resonances extracted in this work provide the first prediction of their pole
properties.

In contrast to the $N(2600)I_{1,11}$ and $N(2700)K_{1,13}$ resonances, 
the extracted properties of the $\Delta(2420)H_{3,11}$ resonance are in a good
agreement with the PDG values. For the other two high-spin $\Delta$ resonances,
i.e., $\Delta(2750)I_{3,13}$ and $\Delta(2950)K_{3,15}$ states, the PDG does not have 
data for comparison. For both $\Delta(2750)I_{3,13}$ and $\Delta(2950)K_{3,15}$
resonances the extracted masses at the pole position are much smaller than those
of the Breit-Wigner. Since the resonances
affect the cross section and other polarization observables only at high energies, 
i.e., $W\gtrsim 2.7$ GeV, a study devoted for high energy photoproduction would be 
very relevant to this end. This would be also in line with the 12 GeV JLab experiments
that are currently in progress. 

In Tables~\ref{tab:pole} and \ref{tab:breit-wigner} 
we have also included the error bars both from PDG and the 
present work. Note that the previous works did not report the uncertainties in the 
extracted resonance masses and widths. The quoted error bars of the present work 
originate from the {\small CERN-MINUIT} output used for the fitting process, where 
we employed the {\small MIGRAD} minimizer that produces both estimated error bars
of the fitted parameters and error matrix \cite{James:1975dr}. 

The error bars obtained from the {\small MINUIT} indicate the flexibility of the model 
to the variation of the resonance masses and widths to reproduce the data, i.e., 
the larger the error bars the more flexible the model.
As a consequence, in the energy region where precise experimental data are abundantly 
available the error bars are forced to be small. Thus, we might expect that the error
bars are large below and near the threshold region and for $W \gtrsim 2.4$ GeV (see
Fig.~\ref{fig:data_dist}). This is proven in 
Table~\ref{tab:pole}, where we can see that relatively larger error bars are obtained
in Model A ($K\Lambda$ photoproduction with $W^{\rm thr.}\approx 1610$ MeV) for the 
$N(1440)P_{11}$, $N(1520)D_{13}$, $N(1535)S_{11}$ resonances. In Model C 
($K\Sigma$ photoproduction with $W^{\rm thr.}\approx 1690$ MeV) three more resonances,
i.e., the $N(1650)S_{11}$, $N(1675)D_{15}$, and $N(1680)F_{15}$ states, also exhibit
this phenomenon. Only the $\Delta(1232)P_{33}$ resonance has very small error bars,
since PDG has estimated this resonance with very precise mass and width, whereas during
the fitting process we allowed these parameters to vary within the PDG uncertainties.
In the higher energy region we observe that all baryon resonances
with $m_{R}\gtrsim 2400$ MeV show the relatively larger error bars.

In conclusion, we have observed that the addition of the  high-spin resonances, 
with spins from 11/2 up to 15/2, in our isobar models improves the agreement between
their pole properties extracted in this work and those listed by PDG.

\section{Summary and Conclusion}
\label{sec:summary}
We have derived the spin-11/2, -13/2, and -15/2 resonance amplitudes for 
kaon photoproduction off the nucleon by using the covariant Feynman
diagrammatic technique. 
For this purpose we made use of the consistent interaction Lagrangians 
proposed by Pascalutsa and Vrancx {\it et al.}, as well as the formulation
of spin-(n+1/2) resonance propagator put forwarded by Vrancx {\it et al}. We have
studied the effect of high-spin resonances in the kaon photoproduction 
processes by including the $N(2600)I_{1,11}$ and $N(2700)K_{1,13}$ states in our previous 
model for $K\Lambda$ photoproduction and the 
$N(2600)I_{1,11}$, $N(2700)K_{1,13}$, $\Delta(2420)H_{3,11}$, $\Delta(2750)I_{3,13}$, and $\Delta(2950)K_{3,15}$
states in our previous model developed for $K\Sigma$ photoproduction.
In general, the inclusion of these high-spin resonances improves the
agreement between model calculations and experimental data, which is
indicated by the smaller values of $\chi^2$ in all isospin channels.
The inclusion of the high-spin resonances also helps to overcome the problem
of resonance-dominated model, since the inclusion increases the hadronic
form factor cutoff of the Born terms and, therefore, increases the role of
the Born terms in both $K\Lambda$ and $K\Sigma$ models.
Specifically, in the $K^+\Lambda$ channel the inclusion leads to fewer 
resonance structures in the cross sections and polarization observables. 
The effect is significant in the high energy differential cross section, 
near the resonance masses and at forward direction. Different from the 
$K^+\Lambda$ channel, the effect in the $K^0\Lambda$ channel is more obvious 
and can be observed in both low and high energy regions. Furthermore,
in this channel the effect is found to be large in both forward and backward angles.
In the $K^+\Sigma^0$ channel the effect is only significant in the forward 
region, where a number of resonance structures appear after including the high-spin
resonances. Two of them are important to note here, i.e., the $\Delta(2000)F_{35}$
and $N(2290)G_{19}$ resonances, which are responsible for the second and third
peaks in the $K^+\Sigma^0$ differential cross section. In contrast to the
$K^+\Sigma^0$ channel, the $K^0\Sigma^+$ channel is found to be sensitive
to these high-spin resonances. In this case, the effect can be observed in
the whole energy range covered by experimental data and the whole angular
distribution. The effect is, however, not observed in the $K^+\Sigma^-$ and 
$K^0\Sigma^0$ channels, at least in the whole kinematics where experimental data
are available. Finally, we found that the addition of the high-spin resonances
leads to a better agreement between the extracted resonance properties and 
those listed by PDG. 

\section*{Acknowledgments}

This work has been supported by the PUTI Q1 Grant of Universitas Indonesia, 
under contract No. BA-1080/UN2.RST/PPM.00.03.01/2020.

\appendix

\begin{widetext}
\section{Form Functions $A_i$ for Baryon Resonances with  Spins 11/2, 13/2 and 15/2}
\label{app:form_function}
The extracted form functions $A_i$ given in Eq.~(\ref{inv}) for baryon resonances with 
spin 11/2 read
\begin{eqnarray}
	A_1 &=& \Bigl[\, 14c_1D_1(s\pm m_{N^*}m_\Lambda)+10D_3 \Bigl\{(m_\Lambda^2 c_k -b_pc_\Lambda)\pm m_{N^*}m_\Lambda (\textstyle{\frac{1}{s}}c_kc_\Lambda -b_p) \Bigr\} \,\Bigr]\,G_1 \nonumber\\
	& &  +\textstyle{\frac{1}{2}}(m_p+m_\Lambda)\Bigl[\,\Bigl\{(m_pc_\Lambda-m_\Lambda c_p)\pm m_{N^*}(m_pm_\Lambda -\textstyle{\frac{1}{s}}c_\Lambda c_p)\Bigr\}\Bigl\{28c_1(c_1-b_pc_s)D_4 -\textstyle{\frac{4}{s}}k^2c_2c_pD_5+D_3\Bigr\} \nonumber\\
	& & -b_p(m_\Lambda\pm \textstyle{\frac{1}{s}}m_{N^*}c_\Lambda )D_3
	\pm 7m_{N^*}\Bigl\{(c_1-b_pc_s)D_2 -\textstyle{\frac{4}{s}}k^2c_1c_2c_pD_4\Bigr\} \,\Bigr]\,G_2  +\Bigl[\, \textstyle{\frac{1}{2}}(m_p+m_\Lambda)\Bigl\{(s\pm m_{N^*}m_p)\nonumber\\
	& & \times \bigl\{-7c_sD_2 -4c_1c_2D_4\textstyle{\frac{1}{s}}c_k\bigr\} -\bigl\{(m_pm_\Lambda +11c_\Lambda) \pm m_{N^*}(\textstyle{\frac{1}{s}}m_pc_\Lambda +11m_\Lambda)\bigr\}D_3 \nonumber\\
	& & -\bigl\{(b_pc_\Lambda -m_pm_\Lambda c_k)\pm m_{N^*}(\textstyle{\frac{1}{s}}m_pc_\Lambda c_k -m_\Lambda b_p)\bigr\}(-28c_1c_sD_4-4c_2D_5\textstyle{\frac{1}{s}}c_k)\Bigr\} \nonumber\nonumber\\
	& & +\Bigl\{(6m_pc_\Lambda +6m_\Lambda c_p +m_\Lambda c_k)\pm m_{N^*}(6m_pm_\Lambda+\textstyle{\frac{6}{s}}c_pc_\Lambda +\textstyle{\frac{1}{s}}c_kc_\Lambda)\Bigr\}D_3\mp 7m_{N^*}c_1D_1 \,\Bigr]\,G_3 \, ,\\
	A_2 &=& {\frac{1}{t - m_K^2}}\Bigl\{\Bigl[-20k^2D_3(c_\Lambda \pm m_{N^*}m_\Lambda)\Bigr]\,G_1+ \Bigl[7(-s\pm m_{N^*}m_p) \Bigl\{(b_pc_s-c_1)D_2 +\textstyle{\frac{4}{s}}k^2c_1c_2c_pD_4\Bigr\} \nonumber \\
	& &  -m_pk^2(m_\Lambda\pm\textstyle{\frac{1}{s}}m_{N^*}c_\Lambda) D_3 +4\Bigl\{(m_pm_\Lambda c_k-b_pc_\Lambda)\pm m_{N^*}(\textstyle{\frac{1}{s}}m_pc_\Lambda c_k-m_\Lambda b_p)\Bigr\}\Bigl\{7c_1D_4(b_pc_s-c_1)\nonumber \\
	& &  +\textstyle{\frac{1}{s}}k^2c_2c_pD_5\Bigr\}\,\Bigr]\,G_2 + m_p\Bigl[(m_\Lambda\pm \textstyle{\frac{1}{s}}m_{N^*}c_\Lambda) \Bigl\{D_3k^2 +56c_1c_kD_4(c_1-b_pc_s) -8c_2\textstyle{\frac{1}{s}}k^2c_k c_pD_5\Bigr\} \nonumber \\
	& &  +\Bigl\{(m_pc_\Lambda +m_\Lambda c_p)\pm m_{N^*}(m_pm_\Lambda +\textstyle{\frac{1}{s}}c_\Lambda c_p)\Bigr\} 4k^2(7c_1c_sD_4+\textstyle{\frac{1}{s}}c_2c_kD_5)\nonumber \\
	& & \pm7 m_{N^*}\Bigl\{D_2(k^2c_s-2b_q)-4c_1c_2k^2D_4(1+\textstyle{\frac{1}{s}}c_p)\Bigr\}\, \Bigr]\,G_3\Bigr\} \, ,\\
	A_3 &=& \Bigl[\, 10c_kD_3(m_\Lambda \pm\textstyle{\frac{1}{s}}m_{N^*}c_\Lambda)\pm 14m_{N^*}c_1D_1 \,\Bigr]\,G_1 +\textstyle{\frac{1}{2}}\Bigl[\, \Bigl\{(m_pc_\Lambda -m_\Lambda c_p)\pm m_{N^*}(m_pm_\Lambda -\textstyle{\frac{1}{s}}c_\Lambda c_p)\Bigr\} \nonumber \\
	& & \times \Bigl\{28c_1D_4\bigl(c_1+b_p(1+\textstyle{\frac{1}{s}}c_\Lambda)\bigr)
	-\textstyle{\frac{4}{s}}k^2c_2c_pD_5 + D_3\Bigr\}-b_p(m_\Lambda\pm \textstyle{\frac{1}{s}}m_{N^*}c_\Lambda )D_3\nonumber \\
	& & \pm 7m_{N^*}\Bigl\{\bigl(c_1+b_p(1+\textstyle{\frac{1}{s}}c_\Lambda)\bigr)\nonumber D_2 -\textstyle{\frac{4}{s}}k^2c_1c_2c_pD_4\Bigr\} \,\Bigr]\,G_2   +\textstyle{\frac{1}{2}}\Bigl[\, 7(s\pm m_{N^*}m_p)\Bigl\{(1+\textstyle{\frac{1}{s}}c_\Lambda)D_2 -4c_1c_2\textstyle{\frac{1}{s}}c_kD_4\Bigr\} \nonumber \\
	& & 	-\Bigl\{(m_pm_\Lambda +11c_\Lambda)\pm m_{N^*}(\textstyle{\frac{1}{s}}m_pc_\Lambda +11m_\Lambda)\Bigr\}
	D_3 -\Bigl\{(b_pc_\Lambda -m_pm_\Lambda c_k)\pm m_{N^*}(\textstyle{\frac{1}{s}}m_pc_\Lambda c_k -m_\Lambda b_p)\Bigr\}\nonumber \\
	& & \times \Bigl\{28c_1D_4(1+\textstyle{\frac{1}{s}}c_\Lambda)-4c_2D_5\textstyle{\frac{1}{s}}c_k\Bigr\} \,\Bigr]\,G_3 \, ,\\
	A_4 &=& \Bigl[\, 10c_kD_3(m_\Lambda \pm\textstyle{\frac{1}{s}}m_{N^*}c_\Lambda)\pm 14m_{N^*}c_1D_2 \,\Bigr]\,G_1 +\textstyle{\frac{1}{2}}\Bigl[\, \Bigl\{(m_pc_\Lambda -m_\Lambda c_p)\pm m_{N^*}(m_pm_\Lambda -\textstyle{\frac{1}{s}}c_\Lambda c_p)\Bigr\}\nonumber \\
	& &\times \Bigl\{28c_1D_4(c_1-b_pc_s) -\textstyle{\frac{4}{s}}k^2c_2c_pD_5 +D_3\Bigr\}-b_p(m_\Lambda\pm \textstyle{\frac{1}{s}}m_{N^*}c_\Lambda D_3\pm 7m_{N^*}\Bigl\{(c_1-b_pc_s)D_2 \nonumber \\
	& & -\textstyle{\frac{4}{s}}k^2c_1c_2c_pD_4\Bigr\} \,\Bigr]\,G_2 +\textstyle{\frac{1}{2}}\Bigl[\, 7(s\pm m_{N^*}m_p)\Bigl\{-c_sD_2-4c_1c_2D_4\textstyle{\frac{1}{s}}c_k\Bigr\} -\Bigl\{(m_pm_\Lambda +11c_\Lambda \nonumber \\
	& & \pm m_{N^*}(\textstyle{\frac{1}{s}}m_pc_\Lambda +11m_\Lambda)\Bigr\}D_3-\Bigl\{(b_pc_\Lambda -m_pm_\Lambda c_k)\pm m_{N^*}(\textstyle{\frac{1}{s}}m_pc_\Lambda c_k -m_\Lambda b_p)\Bigr\} \nonumber\\
 && \times\Bigl\{-28c_1c_sD_4-4c_2D_5\textstyle{\frac{1}{s}}c_k\Bigr\} \,\Bigr]\,G_3 \, ,\\
	A_5 &=& {\frac{1}{t - m_K^2}}\Bigl\{\Bigl[\,10c_5D_3(c_\Lambda \pm m_{N^*}m_\Lambda)\Bigr]\,G_1 + \textstyle{\frac{1}{2}}\Bigl[7(-s \pm m_{N^*}m_p)\Bigl\{D_2(b_p+b_\Lambda -\textstyle{\frac{1}{s}}c_\Lambda c_5)  -\textstyle{\frac{4}{s}}c_1c_2c_5c_pD_4\Bigr\} \nonumber\\
	& &  +m_pc_5(m_\Lambda \pm \textstyle{\frac{1}{s}} m_{N^*}c_\Lambda)D_3 +4\Bigl\{(m_pm_\Lambda c_k-b_pc_\Lambda)\pm m_{N^*}(\textstyle{\frac{1}{s}}m_pc_\Lambda c_k -m_\Lambda b_p)  \Bigr\}\Bigl\{7c_1D_4(b_p+b_\Lambda -\textstyle{\frac{1}{s}}c_\Lambda c_5) \nonumber\\
	& & -\textstyle{\frac{1}{s}}c_2c_5c_pD_5\Bigr\}  \Bigr]\,G_2 + m_p\Bigl[\, \textstyle-{\frac{1}{2}}(m_\Lambda \pm \textstyle{\frac{1}{s}}m_{N^*}c_\Lambda )\Bigl\{c_5\Bigl\{D_3 - \textstyle{\frac{8}{s}}c_2c_kc_pD_5\Bigr\} +56c_1c_kD_4 (b_p+b_\Lambda-\textstyle{\frac{1}{s}}c_\Lambda c_5) \Bigr\} \nonumber\\
	& & -\Bigl\{(m_pc_\Lambda +m_\Lambda c_p)\pm m_{N^*}(m_pm_\Lambda +\textstyle{\frac{1}{s}} c_\Lambda c_p) \Bigr\} 
	\Bigl\{28c_1D_4(4b_\Lambda -k^2-\textstyle{\frac{1}{s}}c_\Lambda c_5)  +\textstyle{\frac{4}{s}}c_2c_5c_kD_5\Bigr\} \nonumber\\
	& &\pm \textstyle{\frac{7}{2}}m_{N^*}\Bigl\{D_2(k^2-2b_q+\textstyle{\frac{1}{s}}c_\Lambda c_5)  +4c_1c_2c_5D_4(1+\textstyle{\frac{1}{s}}c_p)\Bigr\}\, \Bigr]\,G_3\Bigr\} \, ,\\
	A_6 &=& \Bigl[\,10\Bigl\{(m_\Lambda s +m_pc_\Lambda)\pm m_{N^*}(c_\Lambda +m_pm_\Lambda)\Bigr\}D_3\,\Bigr]\,G_1\ + \textstyle{\frac{1}{2}}\Bigl[\Bigl\{(m_pc_\Lambda -m_\Lambda c_p)\pm m_{N^*}(m_pm_\Lambda -\textstyle{\frac{1}{s}}c_\Lambda c_p) \Bigr\}\nonumber \\
	& & \times \Bigl\{D_3 +28c_1D_4(c_1+b_pc_s)  -\textstyle{\frac{4}{s}}c_2c_4c_pD_5\Bigr\} +b_p(m_\Lambda \pm\textstyle{\frac{1}{s}}m_{N^*}c_\Lambda)D_3 \nonumber\\
 && \pm 7m_{N^*}\Bigl\{D_2(c_1+b_pc_s)  -\textstyle{\frac{4}{s}}c_1c_2c_4c_pD_4\Bigr\}\, \Bigr]\,G_2 
+ \textstyle{\frac{1}{2}}m_p\Bigl[7(s \pm m_{N^*}m_p)\Bigl\{D_2c_s   -4c_1c_2D_4(1+\textstyle{\frac{1}{s}}c_p)\Bigr\} \nonumber \\
	& &  +D_3\Bigl\{(m_pm_\Lambda -9c_\Lambda) \pm m_{N^*}(\textstyle{\frac{1}{s}}m_pc_\Lambda -9m_\Lambda)\Bigr\} +4\bigl\{7c_1c_sD_4 +c_2D_5(1+\textstyle{\frac{1}{s}}c_p) \bigr\}\Bigl\{(m_pm_\Lambda c_k- b_pc_\Lambda) \nonumber \\
	& &\pm m_{N^*}(\textstyle{\frac{1}{s}}m_pc_\Lambda c_k -m_\Lambda b_p) \Bigr\}\, \Bigr]\,G_3 \, ,
\end{eqnarray}
whereas the extracted functions $A_i$ for baryon resonances with spin 13/2 read
\begin{eqnarray}
	A_1 &=& \Bigl[\,2D_6(-s\pm m_{N^*}m_\Lambda)+12c_1D_7 \Bigl\{-(m_\Lambda^2 c_k -b_pc_\Lambda)\pm m_{N^*}m_\Lambda (\textstyle{\frac{1}{s}}c_kc_\Lambda -b_p) \Bigr\} \,\Bigr]\,G_1 \nonumber \\
	& & +\textstyle{\frac{1}{2}}(m_p+m_\Lambda)\Bigl[\,\Bigl\{-(m_pc_\Lambda -m_\Lambda c_p) \pm m_{N^*}(m_pm_\Lambda -\textstyle{\frac{1}{s}}c_\Lambda c_p)\Bigr\}\Bigl\{5D_2(c_1-b_pc_s)-\textstyle{\frac{20}{s}}k^2c_1c_2c_pD_4 +c_1D_8\Bigr\}\nonumber \\
	& & -b_p(-m_\Lambda\pm \textstyle{\frac{1}{s}}m_{N^*}c_\Lambda )c_1D_8 \pm m_{N^*}\Bigl\{3c_1(c_1-b_pc_s)D_9 -\textstyle{\frac{5}{s}}k^2c_2c_pD_2\Bigr\} \,\Bigr]\,G_2  +\Bigl[\, \textstyle{\frac{1}{2}}(m_p+m_\Lambda)\nonumber \\
	& & \times \Bigl\{(-s\pm m_{N^*}m_p)\bigl\{-3c_sc_1D_9 -5\textstyle{\frac{1}{s}}c_2c_kD_2\bigr\}-c_1\bigl\{-(m_pm_\Lambda +13c_\Lambda) \pm m_{N^*}(\textstyle{\frac{1}{s}}m_pc_\Lambda+13m_\Lambda)\bigr\}D_8 \nonumber \\
	& & +5\bigl\{-(b_pc_\Lambda -m_pm_\Lambda c_k)\pm m_{N^*}(\textstyle{\frac{1}{s}}m_pc_\Lambda c_k -m_\Lambda b_p)\bigr\}\bigl(c_sc_1^2D_2+\textstyle{\frac{4}{s}}c_1c_2c_kD_4\bigr)\Bigr\} \nonumber \\
	& & +c_1\Bigl\{-(7m_pc_\Lambda  +7m_\Lambda c_p +m_\Lambda c_k)\pm m_{N^*}(7m_pm_\Lambda +\textstyle{\frac{7}{s}}c_pc_\Lambda +\textstyle{\frac{1}{s}}c_kc_\Lambda)\Bigr\}D_8 \mp m_{N^*}D_6 \,\Bigr]\,G_3
\, ,\\
	A_2 &=& {\frac{1}{t - m_K^2}}\Bigl\{\Bigl[-24k^2c_1D_7(-c_\Lambda \pm m_{N^*}m_\Lambda)\Bigr]\,G_1+ \Bigl[(s\pm m_{N^*}m_p)\Bigl\{3c_1(b_pc_s-c_1) D_9 +\textstyle{\frac{5}{s}}k^2c_2c_pD_2\Bigr\} \nonumber \\
	& & -m_pk^2c_1(-m_\Lambda\pm\textstyle{\frac{1}{s}}m_{N^*}c_\Lambda)D_8  +\Bigl\{-(m_pm_\Lambda c_k-b_pc_\Lambda)\pm m_{N^*}(\textstyle{\frac{1}{s}}m_pc_\Lambda c_k-m_\Lambda b_p)\Bigr\}\Bigl\{5D_2(b_pc_s-c_1) \nonumber \\
	& & +\textstyle{\frac{20}{s}}k^2c_1c_2c_pD_4\Bigr\}\,\Bigr]\,G_2+m_p\Bigl[(-m_\Lambda\pm \textstyle{\frac{1}{s}}m_{N^*}c_\Lambda)\Bigl\{k^2c_1D_8  +10c_k(c_1-b_pc_s)D_2 -\textstyle{\frac{40}{s}}k^2c_1c_2c_k c_pD_4\Bigr\}\nonumber \\
	& &  +\Bigl\{-(m_pc_\Lambda +m_\Lambda c_p)\pm m_{N^*}(m_pm_\Lambda +\textstyle{\frac{1}{s}}c_\Lambda c_p)\Bigr\}5k^2\bigl(c_sD_2 +\textstyle{\frac{4}{s}}c_1c_2c_kD_4\bigr)  \nonumber \\
	& & \pm m_{N^*}\Bigl\{3c_1D_9(k^2c_s-2b_q)-5k^2c_2(1+\textstyle{\frac{1}{s}}c_p)D_2\Bigr\}\, \Bigr]\,G_3\Bigr\} \, ,\\
	A_3 &=& \Bigl[\, 12c_kc_1D_7(-m_\Lambda \pm\textstyle{\frac{1}{s}}m_{N^*}c_\Lambda)\pm 2m_{N^*}D_6 \,\Bigr]\,G_1 +\textstyle{\frac{1}{2}}\Bigl[\, \Bigl\{-(m_pc_\Lambda -m_\Lambda c_p)\pm m_{N^*}(m_pm_\Lambda-\textstyle{\frac{1}{s}}c_\Lambda c_p)\Bigr\}\nonumber \\
	& & \times \Bigl\{5D_2 + c_2^2c_3^2)\bigl(c_1+b_p(1+\textstyle{\frac{1}{s}}c_\Lambda)\bigr)-\textstyle{\frac{20}{s}}k^2c_1c_2c_pD_4 +c_1D_2\Bigr\}-b_pc_1(-m_\Lambda\pm \textstyle{\frac{1}{s}}m_{N^*}c_\Lambda )D_8 \nonumber \\
	& &  \pm m_{N^*}\Bigl\{3c_1\bigl(c_1+b_p(1+\textstyle{\frac{1}{s}}c_\Lambda)\bigr)D_9 -\textstyle{\frac{5}{s}}k^2c_2c_pD_2\Bigr\} \,\Bigr]\,G_2  +\textstyle{\frac{1}{2}}\Bigl[\, (-s\pm m_{N^*}m_p)\Bigl\{3c_1(1+\textstyle{\frac{1}{s}}c_\Lambda)D_9 \nonumber \\
	& &  -\textstyle{\frac{5}{s}}c_2c_kD_2\Bigr\} -c_1\Bigl\{-(m_pm_\Lambda +13c_\Lambda)\pm m_{N^*}(\textstyle{\frac{1}{s}}m_pc_\Lambda +13m_\Lambda)\Bigr\}D_8-5\Bigl\{-(b_pc_\Lambda -m_pm_\Lambda c_k)\nonumber\\
	& & \pm m_{N^*}(\textstyle{\frac{1}{s}}m_pc_\Lambda c_k -m_\Lambda b_p)\Bigr\}\Bigl\{(1+\textstyle{\frac{1}{s}}c_\Lambda)D_2-\textstyle{\frac{4}{s}}c_1c_2c_kD_4\Bigr\} \,\Bigr]\,G_3 \, ,\\
	A_4 &=& \Bigl[\, 12c_kc_1D_7(-m_\Lambda \pm\textstyle{\frac{1}{s}}m_{N^*}c_\Lambda)\pm 2m_{N^*}D_6 \,\Bigr]\,G_1 +\textstyle{\frac{1}{2}}\Bigl[\, \Bigl\{-(m_pc_\Lambda -m_\Lambda c_p)\pm m_{N^*}(m_pm_\Lambda -\textstyle{\frac{1}{s}}c_\Lambda c_p)\Bigr\}\nonumber \\
	& & \times \Bigl\{5D_2(c_1-b_pc_s) -\textstyle{\frac{20}{s}}k^2c_1c_2c_pD_4 +c_1D_8\Bigr\}-b_pc_1(-m_\Lambda\pm \textstyle{\frac{1}{s}}m_{N^*}c_\Lambda )D_8 \pm m_{N^*}\Bigl\{3c_1(c_1-b_pc_s)D_9 \nonumber \\
	& &  -\textstyle{\frac{5}{s}}k^2c_2c_pD_2\Bigr\} \,\Bigr]\,G_2 +\textstyle{\frac{1}{2}}\Bigl[\, (-s\pm m_{N^*}m_p)\bigl(-3c_1c_sD_9 -\textstyle{\frac{5}{s}}c_2c_kD_2\bigr) -c_1\Bigl\{-(m_pm_\Lambda +13c_\Lambda)\nonumber \\
	& & \pm m_{N^*}(\textstyle{\frac{1}{s}}m_pc_\Lambda 13m_\Lambda)\Bigr\}D_8 -5\Bigl\{-(b_pc_\Lambda -m_pm_\Lambda c_k)\pm m_{N^*}(\textstyle{\frac{1}{s}}m_pc_\Lambda c_k -m_\Lambda b_p)\Bigr\}\nonumber\\
&& \times
\bigl(-c_sD_2-\textstyle{\frac{4}{s}}c_1c_2c_kD_4\bigr) \,\Bigr]\,G_3 \, ,\\
	A_5 &=& {\frac{1}{t - m_K^2}}\Bigl\{\Bigl[\, 12c_5c_1D_7\bigl\{-c_\Lambda \pm m_{N^*}m_\Lambda\bigr\}\,\Bigr]\,G_1 + \textstyle{\frac{1}{2}}\Bigl[(s \pm m_{N^*}m_p)\Bigl\{3c_1(b_p+b_\Lambda -\textstyle{\frac{1}{s}}c_\Lambda c_5) D_9 -\textstyle{\frac{5}{s}}c_2c_5c_pD_2\Bigr\}\nonumber \\
	& &  +m_pc_5c_1(-m_\Lambda \pm \textstyle{\frac{1}{s}} m_{N^*}c_\Lambda) D_8 +\Bigl\{-(m_pm_\Lambda c_k-b_pc_\Lambda)\pm m_{N^*}(\textstyle{\frac{1}{s}}m_pc_\Lambda c_k -m_\Lambda b_p) \Bigr\}\nonumber \\
	& & \times \Bigl\{3c_1D_9(b_p+b_\Lambda -\textstyle{\frac{1}{s}}c_\Lambda c_5)   -\textstyle{\frac{20}{s}}c_1c_2c_5c_pD_4\Bigr\}   \,\Bigr]\,G_2  + m_p\Bigl[(m_\Lambda\mp \textstyle{\frac{1}{s}}m_{N^*}c_\Lambda )\Bigl\{\textstyle{\frac{1}{2}}c_1c_5D_8 \nonumber \\
	& & +10c_kD_2(b_p+b_\Lambda-\textstyle{\frac{1}{s}}c_\Lambda c_5) +20c_1c_2c_5D_4\textstyle{\frac{1}{s}}c_pc_k \Bigr\} +\textstyle{\frac{5}{2}}\Bigl\{(m_pc_\Lambda +m_\Lambda c_p) \mp m_{N^*}(m_pm_\Lambda+\textstyle{\frac{1}{s}}c_\Lambda c_p)\Bigr\}\nonumber \\
	& & \times \bigl\{(4b_\Lambda -k^2-\textstyle{\frac{1}{s}}c_\Lambda c_5)D_2 +4c_1c_2c_5\textstyle{\frac{1}{s}}c_kD_4\bigr\}\pm \textstyle{\frac{1}{2}}m_{N^*}\Bigl\{3c_1D_9(k^2-2b_q+\textstyle{\frac{1}{s}}c_\Lambda c_5) \nonumber\\ && 
+5c_2c_5D_2(1+\textstyle{\frac{1}{s}}c_p)\Bigr\}\, \Bigr]\,G_3\Bigr\} \, ,\\
	A_6 &=& \Bigl[\,12c_1\bigl\{-(m_\Lambda s+m_pc_\Lambda)\pm m_{N^*}(c_\Lambda +m_pm_\Lambda) \bigr\}D_7\, \Bigr]\,G_1 + \textstyle{\frac{1}{2}}\Bigl[\Bigl\{-(m_pc_\Lambda -m_\Lambda c_p)\pm m_{N^*}(m_pm_\Lambda -\textstyle{\frac{1}{s}}c_\Lambda c_p) \Bigr\}\nonumber \\
	& & \times \Bigl\{c_1D_8 +5D_2(c_1+b_pc_s) -\textstyle{\frac{20}{s}}c_1c_2c_4c_pD_4\Bigr\} +b_p(-m_\Lambda \pm\textstyle{\frac{1}{s}}m_{N^*}c_\Lambda)c_1D_8 \pm m_{N^*}\Bigl\{3c_1(c_1+b_pc_s) D_9 \nonumber \\
	& & -\textstyle{\frac{5}{s}}c_2c_4c_pD_2\Bigr\}\Bigr]\,G_2 +   \textstyle{\frac{1}{2}}m_p\Bigl[(-s \pm m_{N^*}m_p)\Bigl\{3c_1D_9c_s  -5c_2D_2(1+\textstyle{\frac{1}{s}}c_p)\Bigr\} -c_1D_8\Bigl\{(m_pm_\Lambda -11c_\Lambda) \nonumber \\
	& & \mp m_{N^*}(\textstyle{\frac{1}{s}}m_pc_\Lambda -11m_\Lambda) \Bigr\} +\bigl\{5D_2c_s -20c_1c_2D_4(1+\textstyle{\frac{1}{s}}c_p)\bigr\}\nonumber \\ 
&& \times \Bigl\{(b_pc_\Lambda -m_pm_\Lambda c_k)\pm m_{N^*} (\textstyle{\frac{1}{s}}m_pc_\Lambda c_k -m_\Lambda b_p) \Bigr\}\Bigr]\,G_3 \, ,
\end{eqnarray}
where we have defined
\begin{eqnarray}
	D_1 &=& 33c_1^4+30c_1^2c_2c_3+11c_2^2c_3^2 \, ,\\
	D_2 &=& 33c_1^4+18c_1^2c_2c_3+c_2^2c_3^2 \, , \\
	D_3 &=& 21c_1^4+14c_1^2c_2c_3+c_2^2c_3^2   \, , \\
	D_4 &=& 3c_1^2 + c_2c_3  \, , \\
	D_5 &=& 7c_1^2 + c_2c_3 \, , \\
	D_6 &=& 429c_1^6 + 5(99c_1^4c_2c_3 + 27c_1^2c_2^2c_3^2 + c_2^3c_3^3) \, , \\
	D_7 &=& 33c_1^4 + 45c_1^2c_2c_3 + 5c_2^2c_3^2 \, , \\
	D_8 &=& 33c_1^4+120c_1^2c_2c_3+5c_2^2c_3^2  \, , \\
	D_9 &=& 143c_1^4+110c_1^2c_2c_3+15c_2^2c_3^2 \, ,
\end{eqnarray}
while $c_1,c_2$, $c_3$ and other parameters are defined in Eq.~(\ref{eq:c_defined}).

Finally, the extracted functions $A_i$ for baryon resonances with spin 15/2 read
\begin{eqnarray}
	A_1 &=& \Bigl[\, \Bigl( 12879c_1^7+126c_1F_6\Bigr)(s\pm m_{N^*}m_\Lambda)+\Bigl(14F_3+60c_2^3c_3^3\Bigr) \Bigl\{(m_\Lambda^2 c_k -b_pc_\Lambda)\pm m_{N^*}m_\Lambda (\textstyle{\frac{1}{s}}c_kc_\Lambda -b_p) \Bigr\} \,\Bigr]\,G_1 \nonumber \\
	& &  +\textstyle{\frac{1}{2}}(m_p+m_\Lambda)\Bigl[\,\Bigl\{(m_pc_\Lambda -m_\Lambda c_p)\pm m_{N^*}(m_pm_\Lambda -\textstyle{\frac{1}{s}}c_\Lambda c_p)\Bigr\}\Bigl\{F_4(c_1-b_pc_s)-\textstyle{\frac{1}{s}}k^2F_5c_p +  F_3 \Bigr\} \nonumber \\
	& & -b_p(m_\Lambda\pm \textstyle{\frac{1}{s}}m_{N^*}c_\Lambda )F_3\pm m_{N^*}\Bigl\{(c_1-b_pc_s)F_1 -\textstyle{\frac{1}{s}}k^2F_2c_p\Bigr\} \,\Bigr]\,G_2 \nonumber \\
	& & +\Bigl[\, \textstyle{\frac{1}{2}}(m_p+m_\Lambda)\Bigl\{(s\pm m_{N^*}m_p)\bigl\{-c_sF_1 -\textstyle{\frac{1}{s}}c_kF_2\bigr\} -\bigl\{(m_pm_\Lambda +15c_\Lambda)\pm m_{N^*}(\textstyle{\frac{1}{s}}m_pc_\Lambda +15m_\Lambda)\bigr\}F_3 \nonumber \\
	& & -\bigl\{(b_pc_\Lambda -m_pm_\Lambda c_k)\pm m_{N^*}(\textstyle{\frac{1}{s}}m_pc_\Lambda c_k -m_\Lambda b_p)\bigr\}(-F_4c_s-\textstyle{\frac{1}{s}}c_kF_5)\Bigr\} \nonumber\\
	& & +\Bigl\{(8m_pc_\Lambda +8m_\Lambda c_p +m_\Lambda c_k)\pm m_{N^*}(8m_pm_\Lambda+\textstyle{\frac{8}{s}}c_pc_\Lambda +\textstyle{\frac{1}{s}}c_kc_\Lambda)\Bigr\}F_3 \mp m_{N^*}c_1F_1 \,\Bigr]\,G_3 \, ,\\
	A_2 &=& {\frac{1}{t - m_K^2}}\Bigl\{\Bigl[-4k^2\Bigl(7F_3
	+30c_2^3c_3^3\Bigr)(c_\Lambda \pm m_{N^*}m_\Lambda)\Bigr]\,G_1+ \Bigl[(-s\pm m_{N^*}m_p)\Bigl\{(b_pc_s-c_1)F_1 +\textstyle{\frac{1}{s}}k^2c_pF_2\Bigr\} \nonumber \\
	& &  -m_pk^2(m_\Lambda\pm\textstyle{\frac{1}{s}}m_{N^*}c_\Lambda)  F_3 +\Bigl\{(m_pm_\Lambda c_k-b_pc_\Lambda)\pm m_{N^*}(\textstyle{\frac{1}{s}}m_pc_\Lambda c_k-m_\Lambda b_p)\Bigr\} \nonumber\\ && \times 
\Bigl\{F_4(b_pc_s-c_1) + \textstyle{\frac{1}{s}}k^2c_pF_5\Bigr\}\,\Bigr]\,G_2 
+m_p\Bigl[(m_\Lambda\pm \textstyle{\frac{1}{s}}m_{N^*}c_\Lambda)\Bigl\{F_3k^2 +2c_kF_4(c_1-b_pc_s) 
\nonumber \\
	& &  -\textstyle{\frac{1}{s}}k^22c_kc_pF_5\Bigr\} +\Bigl\{(m_pc_\Lambda +m_\Lambda c_p)\pm m_{N^*}(m_pm_\Lambda +\textstyle{\frac{1}{s}}c_\Lambda c_p)\Bigr\} k^2(c_sF_4 +\textstyle{\frac{1}{s}}c_kF_5) 
\nonumber\\
	& & \pm m_{N^*}\Bigl\{F_1(k^2c_s-2b_q)-F_3k^2(1+\textstyle{\frac{1}{s}}c_p)\Bigr\}\, \Bigr]\,G_3\Bigr\} \, ,\\
	A_3 &=& \Bigl[\, 2c_k\Bigl(7F_3
	+30c_2^3c_3^3\Bigr)(m_\Lambda \pm\textstyle{\frac{1}{s}}m_{N^*}c_\Lambda)\pm 2m_{N^*}\bigl(6435c_1^7+63c_1F_6\bigr) \,\Bigr]\,G_1 \nonumber \\
	& &  +\textstyle{\frac{1}{2}}\Bigl[\, \Bigl\{(m_pc_\Lambda -m_\Lambda c_p)\pm m_{N^*}(m_pm_\Lambda -\textstyle{\frac{1}{s}}c_\Lambda c_p)\Bigr\}\Bigl\{F_4\bigl(c_1+b_p(1+\textstyle{\frac{1}{s}}c_\Lambda)\bigr) -\textstyle{\frac{1}{s}}k^2c_pF_5+F_3\Bigr\}\nonumber \\
	& & -b_p(m_\Lambda\pm \textstyle{\frac{1}{s}}m_{N^*}c_\Lambda )F_3\pm m_{N^*}\Bigl\{\bigl(c_1+b_p(1+\textstyle{\frac{1}{s}}c_\Lambda)\bigr)F_1 -\textstyle{\frac{1}{s}}k^2c_pF_2\Bigr\} \,\Bigr]\,G_2 \nonumber \\
	& &  +\textstyle{\frac{1}{2}}\Bigl[\, (s\pm m_{N^*}m_p)\Bigl\{(1+\textstyle{\frac{1}{s}}c_\Lambda)F_1  -F_2\textstyle{\frac{1}{s}}c_k\Bigr\} -\Bigl\{(m_pm_\Lambda +15c_\Lambda)
\pm m_{N^*}(\textstyle{\frac{1}{s}}m_pc_\Lambda +15m_\Lambda)\Bigr\}F_3\nonumber\\
	& & -\Bigl\{(b_pc_\Lambda -m_pm_\Lambda c_k)\pm m_{N^*}(\textstyle{\frac{1}{s}}m_pc_\Lambda c_k -m_\Lambda b_p)\Bigr\} \Bigl\{F_4(1+\textstyle{\frac{1}{s}}c_\Lambda)-F_5\textstyle{\frac{1}{s}}c_k\Bigr\} \,\Bigr]\,G_3 \, ,\\
	A_4 &=& \Bigl[\, 2c_k\Bigl(7F_3
	+30c_2^3c_3^3\Bigr)(m_\Lambda \pm\textstyle{\frac{1}{s}}m_{N^*}c_\Lambda)\pm 2m_{N^*}\bigl(6435c_1^7+63c_1F_6\bigr) \,\Bigr]\,G_1 \nonumber \\
	& & +\textstyle{\frac{1}{2}}\Bigl[\, \Bigl\{(m_pc_\Lambda -m_\Lambda c_p)\pm m_{N^*}(m_pm_\Lambda -\textstyle{\frac{1}{s}}c_\Lambda c_p)\Bigr\}\Bigl\{F_4(c_1-b_pc_s)  -\textstyle{\frac{1}{s}}k^2c_pF_5+(5c_1^2+c_2c_3)\Bigr\}\nonumber \\
	& & -b_p(m_\Lambda\pm \textstyle{\frac{1}{s}}m_{N^*}c_\Lambda )F_3\pm m_{N^*}\Bigl\{(c_1-b_pc_s) F_1 -\textstyle{\frac{1}{s}}k^2c_pF_2\Bigr\} \,\Bigr]\,G_2 \nonumber \\
	& &  +\textstyle{\frac{1}{2}}\Bigl[\, (s\pm m_{N^*}m_p)\bigl(-c_sF_1 -F_2\textstyle{\frac{1}{s}}c_k\bigr)   -\Bigl\{(m_pm_\Lambda +15c_\Lambda)\pm m_{N^*}(\textstyle{\frac{1}{s}}m_pc_\Lambda +15m_\Lambda)\Bigr\}F_3 \nonumber \\
	& & -\Bigl\{(b_pc_\Lambda -m_pm_\Lambda c_k)
\pm m_{N^*}(\textstyle{\frac{1}{s}}m_pc_\Lambda c_k -m_\Lambda b_p)\Bigr\}\bigl(-c_sF_4-F_5\textstyle{\frac{1}{s}}c_k\bigr) \,\Bigr]\,G_3 \, ,
\end{eqnarray}
\begin{eqnarray}
	A_5 &=& {\frac{1}{t - m_K^2}}\Bigl\{\Bigl[\,2c_5\bigl(7F_3+30c_2^3c_3^3\bigr)(c_\Lambda \pm m_{N^*}m_\Lambda)\Bigr]\,G_1 + \textstyle{\frac{1}{2}}\Bigl[(-s \pm m_{N^*}m_p)\Bigl\{F_1(b_p+b_\Lambda -\textstyle{\frac{1}{s}}c_\Lambda c_5)  -\textstyle{\frac{1}{s}}c_5c_pF_2\Bigr\} \nonumber \\
	& &  +m_pc_5(m_\Lambda \pm \textstyle{\frac{1}{s}} m_{N^*}c_\Lambda)F_3+\Bigl\{(m_pm_\Lambda c_k-b_pc_\Lambda)\pm m_{N^*}(\textstyle{\frac{1}{s}}m_pc_\Lambda c_k -m_\Lambda b_p)  \Bigr\}\Bigl\{F_4(b_p+b_\Lambda -\textstyle{\frac{1}{s}}c_\Lambda c_5) \nonumber \\
	& &  -\textstyle{\frac{1}{s}}c_pc_5F_5\Bigr\}  \Bigr]\,G_2+ m_p\Bigl[\, \textstyle-{\frac{1}{2}}(m_\Lambda \pm \textstyle{\frac{1}{s}}m_{N^*}c_\Lambda )\Bigl\{c_5\bigl(F_3 - \textstyle{\frac{1}{s}}c_pc_kF_5\bigr)
+c_kF_4 \bigl(b_p+b_\Lambda-\textstyle{\frac{1}{s}}c_\Lambda c_5\bigr) \Bigr\}
 \nonumber \\
	& &  -\Bigl\{(m_pc_\Lambda +m_\Lambda c_p)\pm m_{N^*}(m_pm_\Lambda +\textstyle{\frac{1}{s}} c_\Lambda c_p) \Bigr\}\Bigl\{F_4(4b_\Lambda -k^2-\textstyle{\frac{1}{s}}c_\Lambda c_5)
+\textstyle{\frac{1}{s}}c_kc_5F_5\Bigr\} \nonumber \\
	& &  \pm \textstyle{\frac{1}{2}}m_{N^*}\Bigl\{F_1(k^2-2b_q+\textstyle{\frac{1}{s}}c_\Lambda c_5)  +c_5F_2(1+\textstyle{\frac{1}{s}}c_p)\Bigr\}\, \Bigr]\,G_3\Bigr\} \, ,\\
	A_6 &=& \Bigl[\,2\Bigl\{(m_\Lambda s +m_pc_\Lambda)\pm m_{N^*}(c_\Lambda +m_pm_\Lambda)\Bigr\}\bigl
	(7F_3+30c_2^3c_3^3\bigr)\,\Bigr]\,G_1\  \nonumber \\
	& & + \textstyle{\frac{1}{2}}\Bigl[\Bigl\{(m_pc_\Lambda -m_\Lambda c_p)
\pm m_{N^*}(m_pm_\Lambda -\textstyle{\frac{1}{s}}c_\Lambda c_p) \Bigr\}
\Bigl\{F_3  +F_4(c_1+b_pc_s) -\textstyle{\frac{1}{s}}c_pc_4F_5\Bigr\} \nonumber\\ && 
+b_p(m_\Lambda \pm\textstyle{\frac{1}{s}}m_{N^*}c_\Lambda) F_3
\pm m_{N^*}\Bigl\{F_1(c_1+b_pc_s) -\textstyle{\frac{1}{s}}c_4c_pF_2
	\Bigr\}\, \Bigr]\,G_2 \nonumber \\
	& & + \textstyle{\frac{1}{2}}m_p\Bigl[(s \pm m_{N^*}m_p)\Bigl\{F_1c_s  -F_2(1+\textstyle{\frac{1}{s}}c_p)\Bigr\}  +F_3\Bigl\{(m_pm_\Lambda -11c_\Lambda)\pm m_{N^*}(\textstyle{\frac{1}{s}}m_pc_\Lambda -11m_\Lambda)\Bigr\}  \nonumber \\
	& & +\bigl\{c_sF_4 +F_5(1+\textstyle{\frac{1}{s}}c_p) \bigr\}\Bigl\{(m_pm_\Lambda c_k- b_pc_\Lambda) \pm m_{N^*}(\textstyle{\frac{1}{s}}m_pc_\Lambda c_k -m_\Lambda b_p) \Bigr\}\, \Bigr]\,G_3 \, ,
\end{eqnarray}
where in terms of $c_1,c_2$, and $c_3$ defined in Eq.~(\ref{eq:c_defined}) we have used
\begin{eqnarray}
	F_1 &=& 45(143c_1^6+143c_1^4c_2c_3+33c_1^2c_2^2c_3^2 + c_2^3c_3^3)  \, , \\
	F_2 &=& 9c_1c_2(286c_1^4 + 220c_1^2c_2c_3+30c_2^2c_3^2)  \, , \\
	F_3 &=& 429c_1^6 + 495c_1^4c_2c_3+135c_1^2c_2^2c_3^2  \, , \\
	F_4 &=& 18c_1(143c_1^4 + 110c_1^2c_2c_3 + 15c_2^2c_3^2)  \, , \\
	F_5 &=& 30c_2(33c_1^4 +18c_1^2c_2c_3 + c_2^2c_3^2)  \, , \\
	F_6 &=& 143c_1^4c_2c_3 + 55c_1^2c_2^2c_3^2 +5c_2^3c_3^3  \, .
\end{eqnarray}
\end{widetext}

\end{document}